\documentclass[tighten,twocolumn]{aastex62}

\usepackage{longtable}
\usepackage{afterpage}

\usepackage{graphicx}
\usepackage{rotating}
\usepackage{float}

\usepackage{epstopdf}

\usepackage{enumitem}

\begin{document}

\shorttitle{Magnetic interactions in the solar atmosphere observed by \textit{IRIS}. II.}
\shortauthors{Guglielmino et al.}

\title{\textit{IRIS} observations of magnetic interactions in the solar atmosphere\\between pre-existing and emerging magnetic fields. II. UV emission properties.}

\author{Salvo L. Guglielmino}
\affiliation{Dipartimento di Fisica e Astronomia ``Ettore Majorana'' -- Sezione Astrofisica, Universit\`{a} degli Studi di Catania, Via S.~Sofia 78, 95123 Catania, Italy}

\author{Peter R. Young}
\affiliation{Code 671, NASA Goddard Space Flight Center, Greenbelt, MD 20771, USA}
\affiliation{College of Science, George Mason University, Fairfax, VA 22030, USA}
\affiliation{Northumbria University, Newcastle upon Tyne, NE1 8ST, UK}

\author{Francesca Zuccarello}
\affiliation{Dipartimento di Fisica e Astronomia - Sezione Astrofisica, Universit\`{a} degli Studi di Catania,
	Via S.~Sofia 78, 95123 Catania, Italy}
		 			 
\correspondingauthor{Salvo L. Guglielmino}
\email{salvatore.guglielmino@inaf.it}

\begin{abstract}

Multi-wavelength ultraviolet (UV) observations by the \textit{IRIS} satellite in active region NOAA 12529 have recently pointed out the presence of long-lasting brightenings, akin to UV bursts, and simultaneous plasma ejections occurring in the upper chromosphere and transition region during secondary flux emergence. These signatures have been interpreted as evidence of small-scale, recurrent magnetic reconnection episodes between the emerging flux region (EFR) and the pre-existing plage field. Here, we characterize the UV emission of these strong, intermittent brightenings and we study the surge activity above the chromospheric arch filament system (AFS) overlying the EFR. 
We analyze the surges and the cospatial brightenings observed at different wavelengths. We find an asymmetry in the emission between the blue and red wings of the \ion{Si}{4} 1402~\AA{} and \ion{Mg}{2}~k 2796.3~\AA{} lines, which clearly outlines the dynamics of the structures above the AFS that form during the small-scale eruptive phenomena. We also detect a correlation between the Doppler velocity and skewness of the \ion{Si}{4} 1394~\AA{} and 1402~\AA{} line profiles in the UV burst pixels. Finally, we show that genuine emission in the \ion{Fe}{12} 1349.4~\AA{} line is cospatial to the \ion{Si}{4} brightenings. This definitely reveals a pure coronal counterpart to the reconnection event. 

\end{abstract}




\section{Introduction}

Solar observations performed by the \textit{Interface Region Imaging Spectrograph} \citep[\textit{IRIS},][]{DePontieu:14} satellite revealed a plethora of small-scale energy release episodes occurring in the low atmosphere and having an impact on the upper chromosphere and the transition region (TR). This atmospheric layer represents the chromosphere-corona transition, with a steep gradient of temperature that increases from chromospheric values ($\approx 10^4\,\mathrm{K}$) up to coronal values of $1 - 2$~MK in about 2000~km.

Among these energetic phenomena, \citet{Peter:14} pointed out the presence of \textit{IRIS} bombs: that is, intense, small-scale, short-lived brightenings seen in ultraviolet (UV) images. The properties of these transient events, also called UV bursts, have been recently inventorized by \citet{Young:18}, taking advantage of a number of observational studies that focused on this subject \citep[e.g.,][]{Peter:14,Gupta:15,Kim:15,Vissers:15,Grubecka:16,Tian:16,Chitta:17,Hong:17,Luc:17,Zhao:17,Tian:18}. 
UV bursts appear as compact bright grains that exhibit a factor of $100 - 1000$ increase in intensity in UV lines and often show strong line broadening indicating plasma flow ejections of $\approx 200 \,\mathrm{km\,s}^{-1}$. They occur on spatial scales of $\approx 500 - 1000$~km and are observed for a short time ($\sim 5$ minutes), although a larger duration has been also reported: in this case UV bursts appear as a sequence of intermittent, repetitive flarings. Most of them are associated with small-scale, canceling opposite-polarity magnetic flux patches in the photosphere, such as those observed in emerging flux regions (EFRs) embedded in magnetized environments, and share some resemblance to similar phenomena observed at optical wavelengths, like Ellerman bombs \citep{Vissers:15,Tian:16,Libbrecht:17,Nelson:17,Luc:17}. They probably arise as a consequence of small-scale magnetic reconnection occurring in the low atmosphere, at photospheric and/or chromospheric heights. Remarkably, observations of coronal signatures of these events are rare.

The idea that reconnection may occur when an EFR interacts with the pre-existing ambient field, releasing enough energy to heat the solar atmosphere and drive high-temperature plasma flows, was proposed by \citet{Heyvaerts:77} and developed by \citet{Shibata:89,Yokoyama:95,Yokoyama:96}. It was confirmed by a long series of observational studies and numerical simulations \citep[see, e.g.,][and reference therein]{Guglielmino:12,Cheung:14}. Indeed, detailed observations of small-scale EFRs and of their chromospheric and coronal response, carried out in recent years with increasing spatial resolution have reported a similar scenario \citep[e.g.][]{Guglielmino:08,Guglielmino:10,Santiago:12,Ortiz:14,Jaime:15,Centeno:17}. These studies show that regions characterized by small-scale transient brightenings and jet-like ejections all exhibit the presence of opposite magnetic polarities that come into contact and/or cancel with each other at the photospheric level \citep[see also][]{Shimizu:15}. Simulations indicate that the relative orientation between the two interacting flux systems rules the dynamics and the energetics of the process \citep[e.g.,][]{Galsgaard:05,Galsgaard:07}, while the pre-existing field acts as a guide for plasma ejections \citep[e.g.,][]{Fernando:08,David:15}. However, in observations at very high resolution there is still a lack of connectivity between all the layers of the solar atmosphere, from the photosphere to the corona. In this perspective, recent studies have benefited from the \textit{IRIS} capabilities in order to have simultaneous multi-wavelength observations during flux emergence episodes in the TR as well \citep{Santiago:14,Jiang:15,Ortiz:16,Toriumi:17,Tian:18}.

In the previous paper of this series \citep[][henceforth Paper~I]{Guglielmino:18}, we have described the overall evolution in the solar atmosphere of a flux emergence episode occurring within the plage of active region (AR) NOAA~12529. Analyzing simultaneous observations by the \textit{IRIS}, \textit{Hinode}, and \textit{Solar Dynamics Observatory} \citep[SDO,][]{Pesnell:12} satellites during the emergence phase of the EFR, we have determined its photospheric configuration until its full development, together with its chromospheric and coronal response. 

Using multi-wavelength observations in the UV and EUV, we have identified recurrent intense brightenings in the EFR site, which have been recognized as UV bursts, with counterparts in all EUV passbands. These are localized near the contact region between the negative emerging flux and the positive ambient field in the photosphere. In addition, plasma ejections at chromospheric and coronal levels have been found close to the observed brightness enhancements. The analysis of the \textit{IRIS} line profiles has suggested that heating of dense plasma occurs in the low solar atmosphere. Simultaneously, ejections of bi-directional flows with speed up to $\sim 100 \,\mathrm{km\,s}^{-1}$ are observed. We have compared these observational signatures with realistic 2.5 MHD numerical simulations of flux emergence that show similar features \citep{Nobrega:16,Nobrega:17}. This comparison that has been carried out in Paper~I has pointed out that intensity enhancements and plasma ejections are triggered by several long-lasting, small-scale magnetic reconnection episodes, which lead to flux cancelation at the photospheric level. In order to explain the observed coronal counterparts, it has been also proposed that magnetic reconnection takes place higher in the atmosphere than usually found in UV bursts.

Here, we investigate the dynamics related to the chromospheric arch filament system \citep[AFS; e.g.,][]{Bruzek:80} overlying the EFR. Moreover, we focus on the analysis of UV emission during the \textit{IRIS} observations, performing a statistical analysis of the profiles relevant to the UV bursts and providing evidence of a pure coronal counterpart of these brightenings. The layout of the paper is organized as follows. In the next Section we describe the observations. In Sect.~3 we analyze the evolution of the UV spectra in the EFR, characterizing their properties. We discuss the implications of our findings in Sect.~4. Finally, Section~5 contains a summary and our conclusions.

\begin{figure*}[t]
	\centering
	\includegraphics[scale=0.365, clip, trim=55 185 180 30]{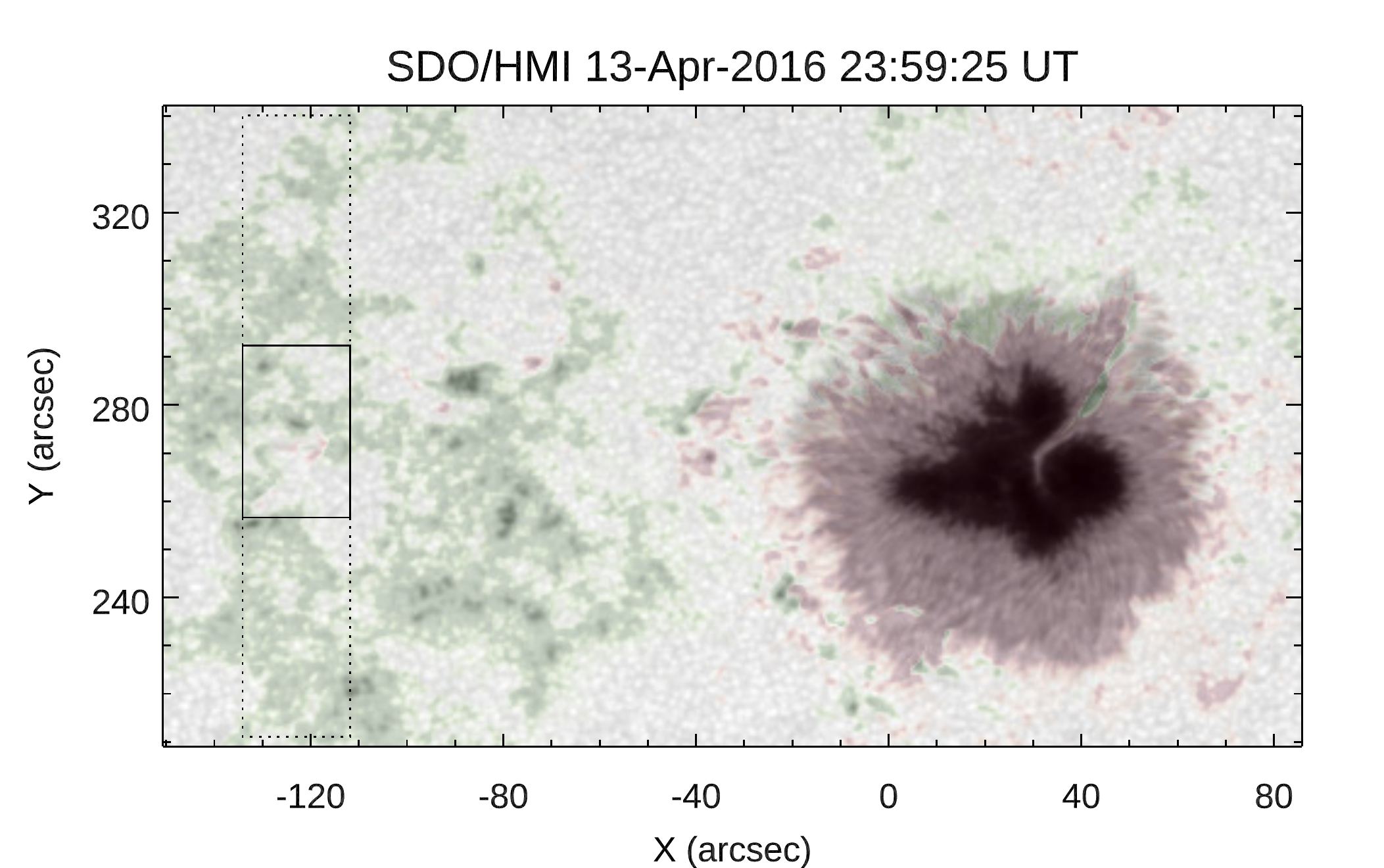}%
	\includegraphics[scale=0.365, clip, trim=25 185 80 30]{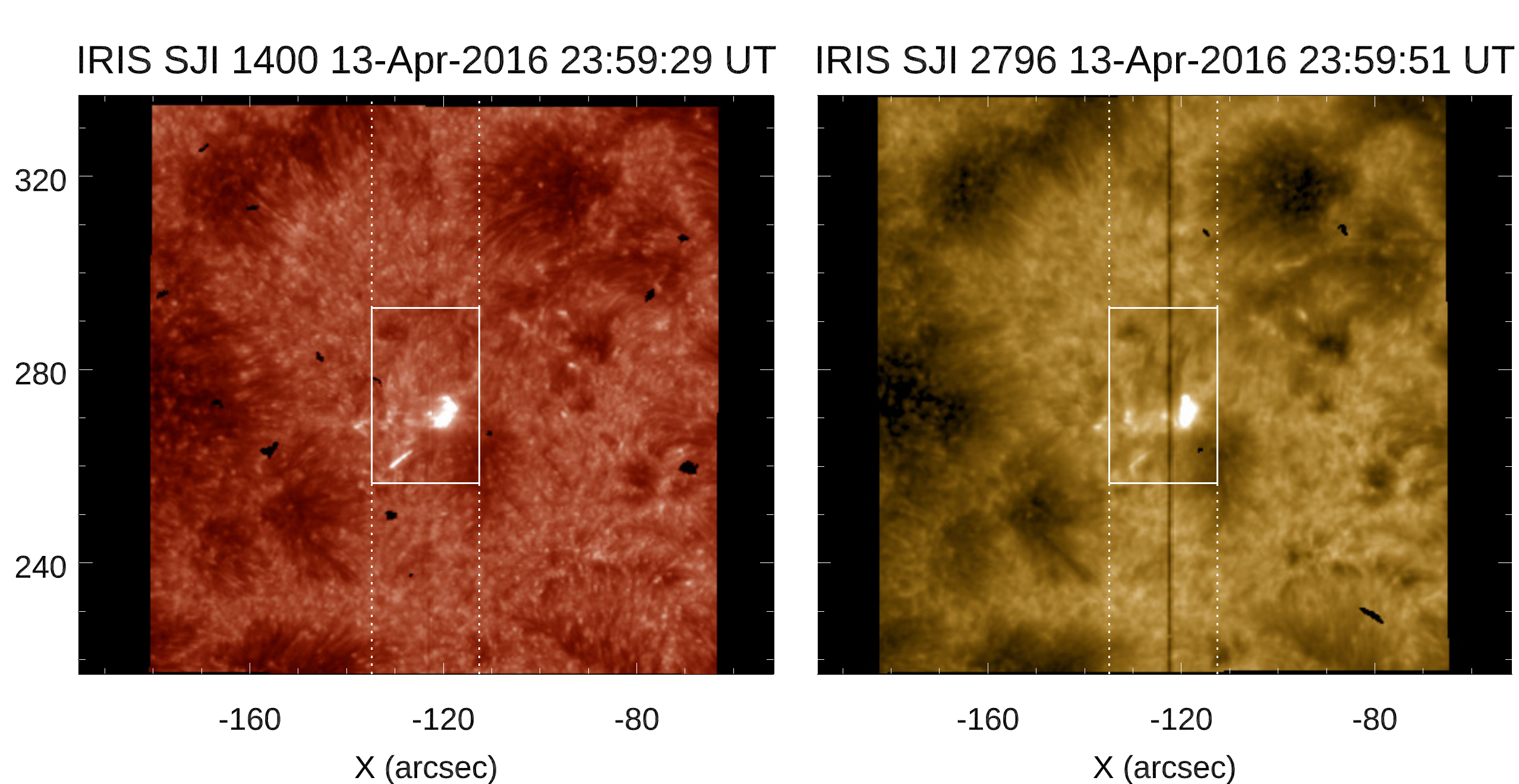}%
	\caption{\textit{Left:} AR NOAA~12529 as seen in the \textit{SDO}/HMI continuum filtergram acquired at midnight between April~13 and~14, with overlaid LOS magnetogram (green/red: positive/negative polarity). The solid box frames the portion of the FoV where the EFR appears. This subFoV is used for the analysis in this paper. The dashed box indicates the area covered by the \textit{IRIS} slit during the six large dense 64-step rasters. \textit{Middle and Right:} \textit{IRIS} SJ 1400~\AA{} and 2796~\AA{} near co-temporal images that outline chromospheric and TR features, relevant to the third raster scan of the analyzed observing sequence. The solid white box marks the common subFoV between the area scanned by the \textit{IRIS} slit and the region where the EFR is observed in the \textit{SDO}/HMI continuum map. The dashed box indicates the area covered by the \textit{IRIS} slit. Here and in the following figures, North is at the top, West is to the right. The axes give the distance from solar disc center. \label{fig_context}}
\end{figure*}

\section{Observations}

AR NOAA~12529 appeared on the solar disc during April 2016, being characterized by a $\beta$-type magnetic configuration (see Figure~\ref{fig_context}). It passed across the central solar meridian between April 13 and 14, when it was located at heliocentric angle $\mu \approx 0.96$ \citep{Guglielmino:17,Guglielmino:18UF}. An EFR was observed emerging at that time, being embedded in the plage field of the following polarity of the AR 
(see the solid box in Figure~\ref{fig_context}). 

Three different data sets were acquired by the \textit{IRIS} satellite during the EFR evolution. The one relevant to our study is an observing sequence acquired between 22:34:43~UT on April~13 and 01:55:29~UT on April~14. The sequence consists of six large dense 64-step raster scans (OBS3610113456). UV spectra were acquired in seven spectral ranges, including \ion{C}{2}~1334.5 and 1335.7~\AA{}, \ion{Si}{4}~1394 and 1402~\AA{}, \ion{Mg}{2}~k~2796.3 and h~2803.5~\AA{} lines and the faint lines around the chromospheric \ion{O}{1} 1355.6~\AA{} line. This spectral interval comprises resonance lines from ions such as the forbidden \ion{Fe}{12} 1349.4~\AA{} line and the very hot \ion{Fe}{21} 1354~\AA{} line, both relevant to the corona. The exposure time was initially 30~s, then on exposure 10 in the second raster the automatic exposure control reduced it to 9~s for the NUV channel and to 18~s for the FUV channels, respectively. These relatively long exposure times ensured a good signal-to-noise ratio for faint lines as well. The sequence had a 0\farcs33 step size and a 31.5~s step cadence, with a pixel size of 0\farcs35 along the \textit{y} direction (spatial binned data). The raster cadence was about 33~min. The field of view (FoV) covered by each scan was $22\farcs2 \times 128\farcs4$, as indicated in Figure~\ref{fig_context} (dashed box). Simultaneously, slit-jaw images (SJIs) were acquired in the 1400 and 2796~\AA{} passbands, relevant to the \ion{Si}{4} 1402~\AA{} and \ion{Mg}{2}~k lines, respectively. These SJIs have a cadence of 63~s for consecutive frames in each passband and cover a FoV of $143\farcs7 \times 128\farcs4$. For further details about this data set and its analysis, we refer the reader to Paper~I.

To determine the context of \textit{IRIS} observations, we used photospheric observations from the \textit{SDO} satellite consisting of full-disk continuum filtergrams and line-of-sight (LOS) magnetograms taken by the Helioseismic and Magnetic Imager \citep[HMI,][]{Scherrer:12} along the \ion{Fe}{1}~6173~\AA{} line, with a spatial resolution of 1\arcsec{} and a cadence of 45~s. Coronal images from the 193~\AA{} filter acquired by the Atmospheric Imaging Assembly \citep[AIA;][]{Lemen:12} were also considered in the present work. These EUV data have an image spatial scale of about 0\farcs6 per pixel and a cadence of 12~s. 

The alignment between \textit{IRIS} and \textit{SDO} observations was obtained by applying cross-correlation techniques with respect to the cospatial subFoV between the \textit{SDO}/HMI continuum filtergrams and each \textit{IRIS} scan. We used the integrated radiance in the \textit{IRIS} 2832~\AA{} band, relevant to the photosphere, where the pores visible in the \textit{SDO}/HMI continuum were taken as fiducial points, considering the pixel scale of the different instruments, as explained in Paper~I. The accuracy of the alignment is $\pm 0\farcs5$, being comparable to the pixel size of \textit{SDO}/HMI data.

\section{Results}

\begin{figure}[b]
	\centering
	\includegraphics[scale=0.575, clip, trim=15 25 115 245]{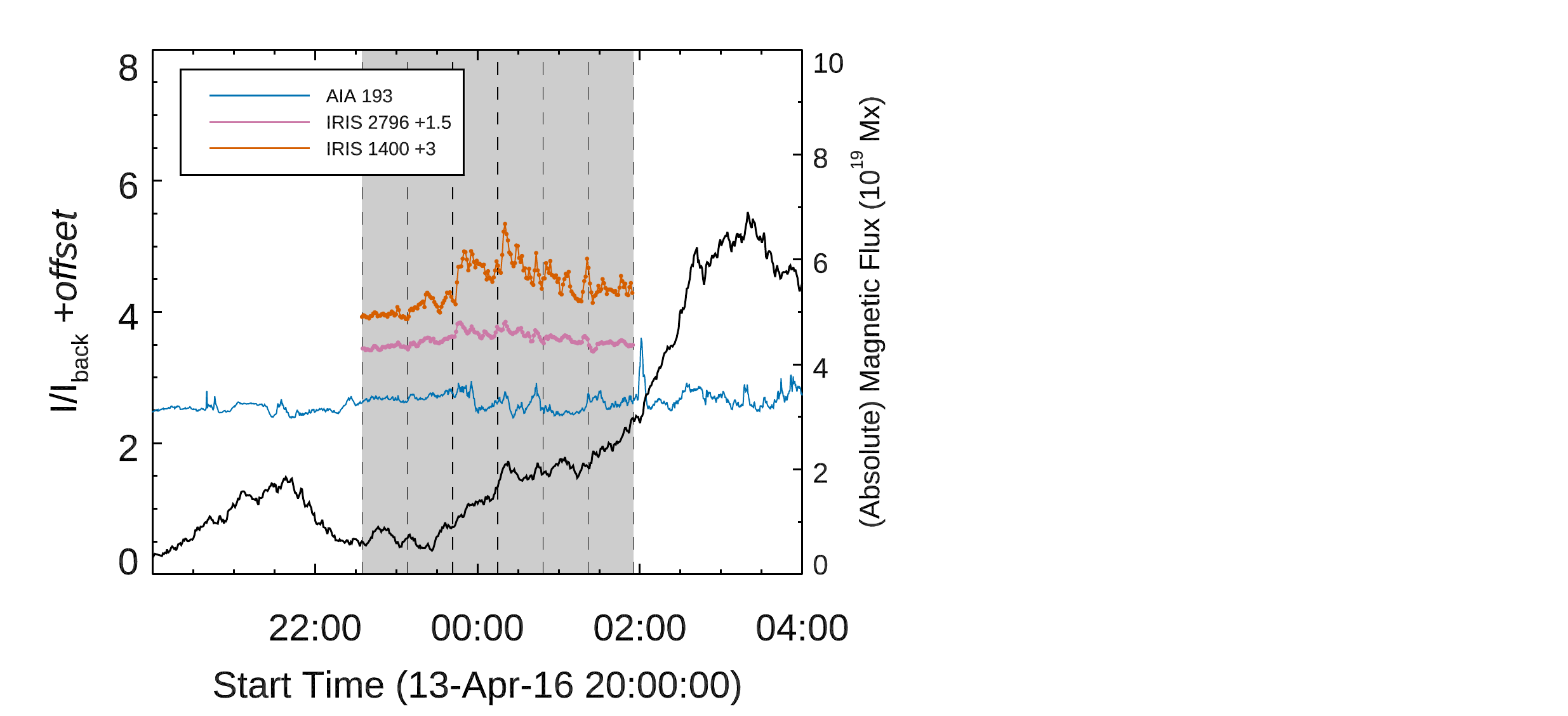}
	\caption{Plot of the flux trend of the negative polarity of the EFR during the observations (solid black line). Light curves in the subFoV shown in Figure~\ref{fig_context} for the \textit{IRIS} UV 1400~\AA{} and 2796~\AA{} SJI passbands and for the \textit{SDO}/AIA EUV 193~\AA{} channel are also plotted (colored lines, according to the legend). Note the offsets of the light curves along the \textit{y} direction, to enhance their visibility. The grey-shaded area indicates the time interval of the \textit{IRIS} rasters. The beginning and end times of each \textit{IRIS} raster are indicated with dashed vertical lines. \label{fig_lightcurve}}
\end{figure}

The EFR analyzed in this study was observed within the subFoV indicated with solid boxes in Figure~\ref{fig_context}. The flux history relevant to the EFR is displayed in the plot of Figure~\ref{fig_lightcurve} (solid black line). This represents the amount of negative flux emerged in the area where the EFR was observed, from 20:00~UT on April~13 until 04:00~UT on April~14, as deduced from \textit{SDO}/HMI LOS magnetograms (see Paper~I). The gray-shaded area indicates the time interval of the \textit{IRIS} sequence, which comprises the emergence phase of the EFR.

In the same graph, we plot the light curve for the \textit{SDO}/AIA 193~\AA{} channel (blue) in the subFoV indicated in Figure~\ref{fig_context}. This light curve shows some brightness enhancements during the flux emergence process. For comparison, we add the light curves for the \textit{IRIS} 2796~\AA{} (magenta) and 1400~\AA{} (orange) SJI passbands in the same subFoV. These make visible the enhanced emission as found in SJIs that exhibits a smoother increase in the 2796~\AA{} passband and a more bursty behaviour in the 1400~\AA{} passband. 

\begin{figure*}[t]
	\centering
	\includegraphics[scale=0.485, clip, trim=35 330 185 80]{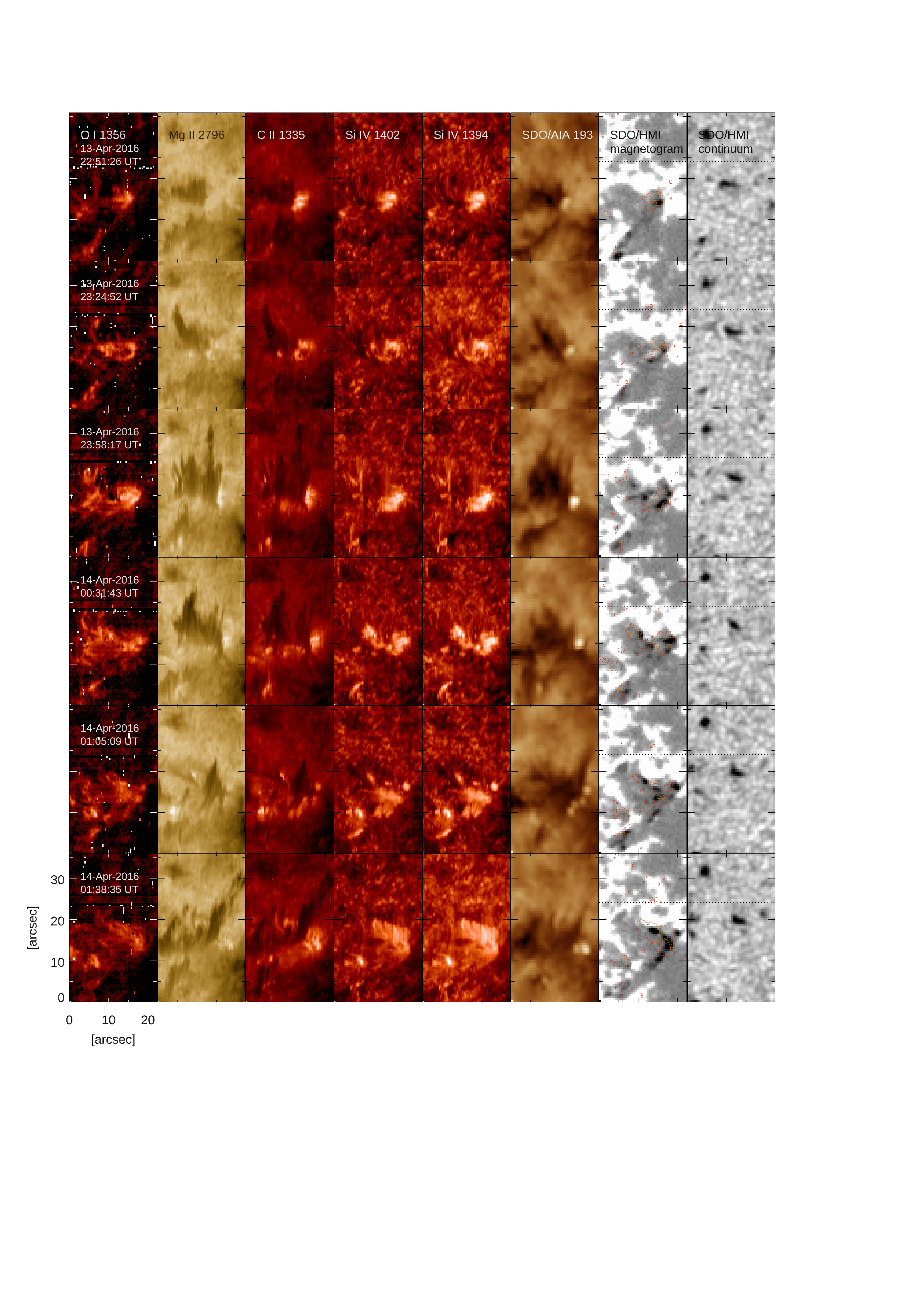}
	\caption{Synoptic view of the evolution of the EFR at different atmospheric layers, during the \textit{IRIS} observing sequence. From left to right, \textit{IRIS} reconstructed maps of the radiance in the \ion{O}{1} 1355.6~\AA{}, \ion{Mg}{2}~k 2796.3~\AA{}, \ion{C}{2}~1335.7~\AA{}, \ion{Si}{4}~1402 and 1394~\AA{} lines. The reference time is that of the halfway raster position of each \textit{IRIS} scan. Moreover, we display the \textit{SDO}/AIA filtergrams in the 193~\AA{} passband and the \textit{SDO}/HMI LOS magnetograms and continuum maps at the closest time to the halfway time of each scan. For comparison, the FoV below the dashed black line in the \textit{SDO}/HMI maps is the same as shown in the synoptic image shown in Figure~6, Paper~I. Contours overplotted on the \textit{SDO}/HMI LOS magnetograms refer to the 60\% (red) and 80\% (orange) of the maximum radiance in the \ion{Si}{4}~1402~\AA{} line, respectively. A logarithmic intensity scaling is used for the \textit{IRIS} reconstructed maps and \textit{SDO}/AIA images. \label{fig_synoptic}}
\end{figure*}


\begin{figure*}[t]
	\centering
	\includegraphics[scale=0.485, clip, trim=35 330 185 80]{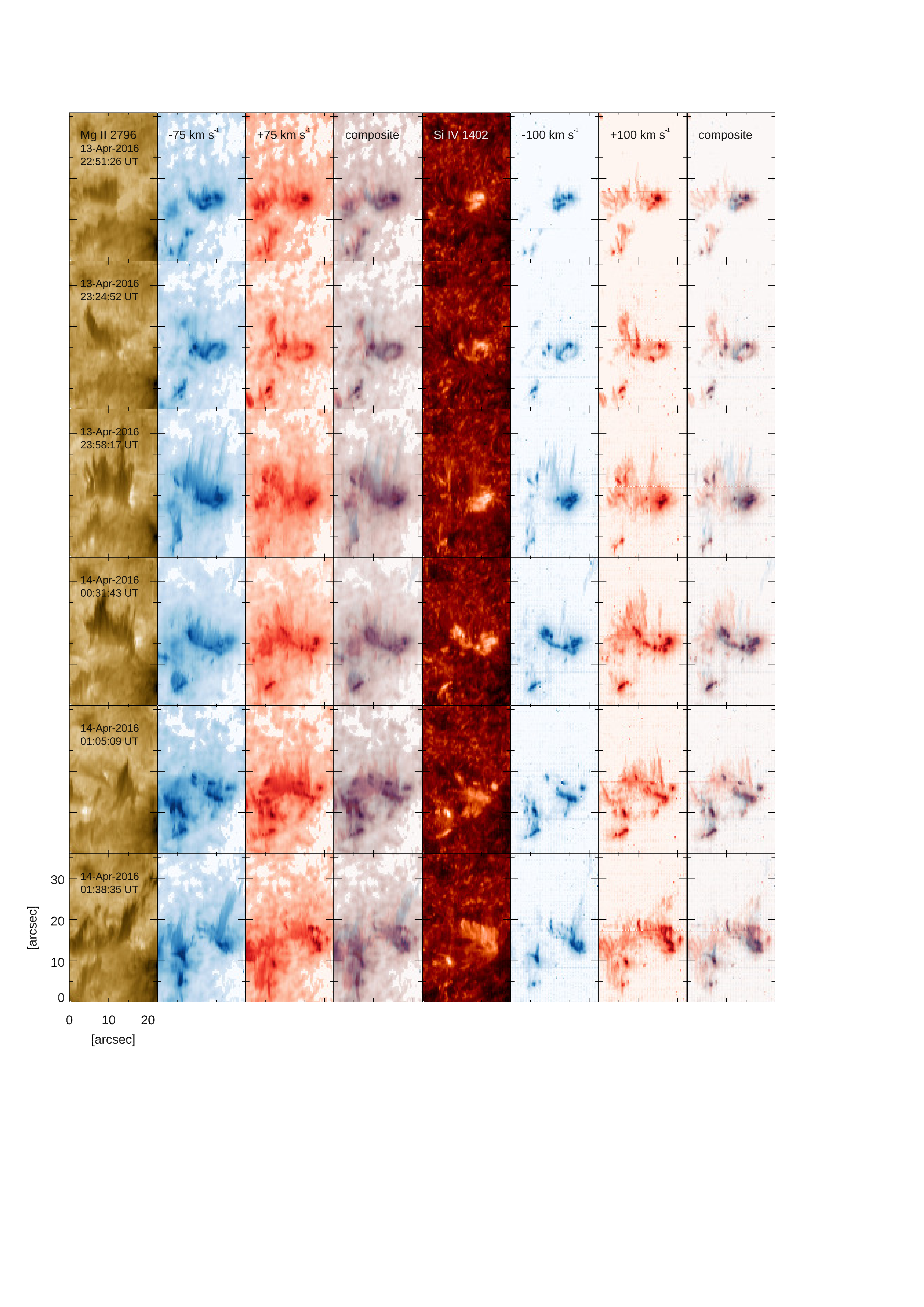}
	\caption{Radiance maps deduced for the line core and for the wings of the chromospheric \ion{Mg}{2}~k 2796.3~\AA{} line (left panels) and the TR \ion{Si}{4}~1402~\AA{} line (right panels). From the left to the right: Radiance map in the line core (first column), in the blue wing at $- 75 \,\mathrm{km\,s}^{-1}$ (second column), in the red wing at $+ 75 \,\mathrm{km\,s}^{-1}$ (third column), and composite image of the blue and red wing radiance maps (fourth column) for the \ion{Mg}{2}~k 2796.3~\AA{} line. Same for the \ion{Si}{4}~1402~\AA{} line (fifth-eighth columns): note that for this line the blue and red wings are evaluated at $\pm 100 \,\mathrm{km\,s}^{-1}$. \label{fig_asymmetry}}
\end{figure*}

Figure~\ref{fig_synoptic} pictures the evolution of the structures seen in the subFoV during the six scans performed with the \textit{IRIS} spectrograph. For \textit{IRIS} lines, we estimated the radiance by averaging the intensity within $\pm 15 \,\mathrm{km\,s}^{-1}$ with respect to the core of each line (laboratory rest wavelength). Then, we derived the reconstructed maps of the radiance. Note that we subtracted the continuum in deriving the radiance for the weak \ion{O}{1} 1355.6~\AA{} line. 

The \ion{O}{1} 1355.6~\AA{} radiance maps (Figure~\ref{fig_synoptic}, first column) image the emission in the middle chromosphere. 
In this layer, we see the evolution of an AFS forming above the EFR. Brightness enhancements are also visible along the AFS, the most conspicuous one being roughly centered at around X=15\arcsec, Y=15\arcsec{} in the third scan (23:58~UT). The \ion{Mg}{2}~k~2796.3 and \ion{C}{2}~133.57~\AA{} maps (Figure~\ref{fig_synoptic}, second and third columns) clearly show the development of dark, elongated structures that we refer to as `threads' in the upper chromosphere. They depart from above the AFS and can be identified with the surges seen in the \textit{IRIS} 2796~\AA{} SJIs (see Figure~8 in Paper I). The surges seen in the third scan (23:58~UT), at the center of the FoV, are particularly developed, as well as the westernmost plasma ejecta observed in the sixth scan (01:38~UT). To the west of these structures, a compact brightening is observed. Such an UV burst has a higher contrast with respect to the background intensity in the \ion{C}{2} line. Some other knots with enhanced emission are located at the base of the surges. The radiance maps deduced for the \ion{Si}{4}~1394 and 1402~\AA{} lines (Figure~\ref{fig_synoptic}, fourth and fifth columns) look very similar, the most evident feature being the UV burst. The fact that the radiance maps of \ion{Si}{4}~1402 and 1394~\AA{} lines are almost identical suggests that the optically thin approximation is valid for the most part. This has been verified using scatter plots of \ion{Si}{4}~1394 vs.~\ion{Si}{4}~1402 intensities reported and discussed in the Appendix (see Figure~\ref{fig_scatter_thinness}). In the first three scans, the UV burst looks compact and slightly more extended than in the \ion{C}{2}~133.57~\AA{} radiance maps. From the fourth scan, the area with enhanced emission comprises both the UV burst and the region beneath the surges, given that the radiation in the \ion{Si}{4} lines is not obscured by their presence. In fact, bright knots that are not visible in the \textit{IRIS} SJ 1400~\AA{} broad-band images are clearly detected, some of which are cospatial to the knots noticed in the \ion{C}{2} line. This area expands in the following scans. Nevertheless, a region with stronger emission to the west of the FoV is still noticeable. 

The cotemporal filtergrams in the \textit{SDO}/AIA 193~\AA{} channel (Figure~\ref{fig_synoptic}, sixth column) show in this subFoV both the surges as dark threads and an EUV brightening as a counterpart of the UV burst seen by \textit{IRIS}. In particular, the EUV brightness enhancements are always cospatial to the UV burst seen in the \ion{C}{2}~1335.7~\AA{} radiance maps, with the same extent. Note that the 193~\AA{} filtergrams do not image exactly the same structures seen in the \textit{IRIS} radiance maps, since the former refer to a single instant of time that is the closest to the halfway time of each \textit{IRIS} raster. The \textit{SDO}/AIA 193 burst could be due to cool, $\log T \left[ \mathrm{K} \right] \sim 5$ emission that contaminates this coronal channel, as discussed by \citet{Winebarger:13} and \citet{Tian:14}. However, we demonstrate below that \ion{Fe}{12} 1349~\AA{} line is detected in the \textit{IRIS} spectra at the burst site, confirming that there is some coronal emission.

Finally, \textit{SDO}/HMI LOS magnetograms and continuum filtergrams are displayed (Figure~\ref{fig_synoptic}, seventh and eighth columns). Again, these refer to the instant of time closest to the halfway time of each \textit{IRIS} raster. The overplotted contours of \ion{Si}{4} 1402~\AA{} radiance essentially indicate that the brightenings seen as UV bursts and EUV enhancements are localized in a region between opposite magnetic polarities at the photospheric levels: the emerging patches with negative polarity of the EFR and pre-existing positive flux concentrations belonging to the plage. The largest pre-existing concentrations correspond to pores in the continuum maps (see also Paper~I).

In Figure~\ref{fig_asymmetry} we display the reconstructed maps of integrated emission in different wavelength ranges along the line profile for both the chromospheric \ion{Mg}{2}~k 2796.3~\AA{} line and the TR \ion{Si}{4}~1402~\AA{} line. The radiance map for each line, as derived in the line core for Figure~\ref{fig_synoptic}, is shown for reference (reference radiance maps, first/fifth column). The other maps image the UV emission in the blue (second/sixth column) and red (third/seventh column) wings of each line. Note that we use $\pm 75 \,\mathrm{km\,s}^{-1}$ for the wings of the \ion{Mg}{2}~k 2796.3~\AA{} line and $\pm 100 \,\mathrm{km\,s}^{-1}$ for the wings of the \ion{Si}{4}~1402~\AA{} line. From the composite maps (fourth/eighth column), it is possible to determine whether there are differences in the location of the emitting regions between the wings. 

\begin{figure*}[!t]
	\centering
	\includegraphics[scale=0.675, clip, trim= 0 10 0 140]{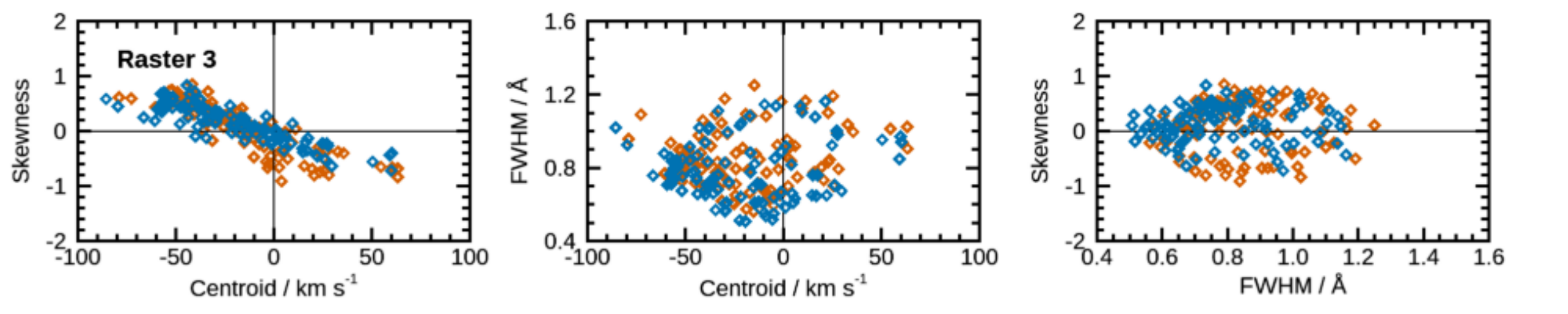}
	\includegraphics[scale=0.675, clip, trim= 0 10 0 140]{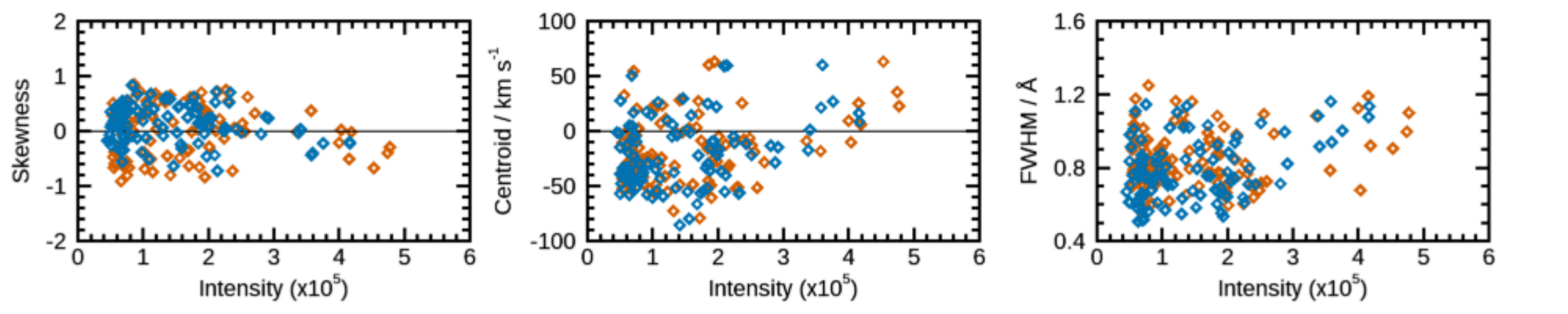}	
	\caption{\textit{Top panels:} Plots of centroid vs.~skewness (left), centroid vs.~FWHM (middle), and FWHM vs.~skewness (right) for both the \ion{Si}{4}~1394~\AA{} (blue symbols) and 1402~\AA{} (red symbols) line profiles for the third \textit{IRIS} raster scan. \textit{Bottom panels:} Same as in the top panels, for the integrated intensity vs.~skewness (left), intensity vs.~centroid (middle), and intensity vs.~FWHM (right). \label{fig_scatter}}
\end{figure*}

Actually, it is not easy to decipher the radiance maps along the blue and red wings of these lines and the composite maps, because the projection effects along the LOS owing to the geometry of the structures may strongly bias out interpretation. Moreover, it has to be remembered that there is a delay of 30~s between an exposure and the following one along the \textit{x} direction. Some observational facts that can be deduced from Figure~\ref{fig_asymmetry} are described as follows.
\begin{itemize}[noitemsep]
	\item Several elongated threads are observed both in blue and red wings of the \ion{Mg}{2}~k and \ion{Si}{4}~1402~\AA{} lines.
	\item Most of these threads take place essentially in the same region occupied by the structures identified as surges in the reference radiance maps.
	\item A number of threads appearing in the blue wing of the \ion{Mg}{2}~k line has a counterpart in the red wing that is shifted toward the West, along the \textit{x} direction. This is more perceivable from the composite maps. Some threads exhibit a similar asymmetry in the \ion{Si}{4}~1402~\AA{} line wings as well.
	\item Some threads observed in the red wing have no counterpart in the blue wing. For instance, this is the case for the thread seen at 23:24~UT centered at X=5\arcsec, Y=20\arcsec{} and that observed at 01:05~UT centered at X=10\arcsec, Y=20\arcsec, both in the \ion{Mg}{2}~k and \ion{Si}{4}~1402~\AA{} lines.
	\item Threads seen in the blue wing of both lines last generally longer than their counterpart seen in the red wings. In general, this is also the case for threads observed in the blue or red wing only.
	\item Most of the threads seen in the blue wings exhibit a south-east--to--north-west orientation, with their top edge more displaced toward the West direction. Conversely, many threads seen in the red wings show a north-east--to--south-west orientation, with their top edge more displaced toward the East direction.
	\item Notably, a couple of prominent threads are seen only in the blue wings. This is the case for the thread seen at 00:31~UT centered at X=20\arcsec, Y=30\arcsec{} (\ion{Si}{4}~1402~\AA{} line) and for that more conspicuous one observed at 01:38~UT centered at X=15\arcsec, Y=20\arcsec, both in the \ion{Mg}{2}~k line and, less clearly, in the \ion{Si}{4}~1402~\AA{} line.
\end{itemize}

In contrast to what is observed in the threads, Figure~\ref{fig_asymmetry} indicates that the emission in the UV bursts does not exhibit large differences between the blue and red wings of the \ion{Mg}{2}~k and \ion{Si}{4}~1402~\AA{} lines. One can notice only a somewhat stronger intensity of the emission in the red wings with respect to that in the blue wings of the lines, whereas the morphology and size of the UV bursts are essentially the same in both wings.

A further information can be obtained from a comparison between Figure~\ref{fig_synoptic} and Figure~\ref{fig_asymmetry}. Two compact bright knots are visible in the radiance maps, being centered at X=5\arcsec, Y=5\arcsec{} and, since 01:05~UT, also at X=5\arcsec, Y=10\arcsec. The kernels of these sites with enhanced brightness are clearly visible in the lines relevant to the upper chromosphere and TR (\ion{Mg}{2}~k, \ion{C}{2}~1335.7~\AA{}, and \ion{Si}{4} lines) and appear having the same aspect and size in both wings of the \ion{Mg}{2}~k and \ion{Si}{4}~1402~\AA{} lines. Threads with a smaller size with respect to those previously discussed are seen departing from these knots, especially in the blue wings of the lines. These threads are indeed cospatial with small surge-like ejections visible in the \ion{Mg}{2}~k and \ion{C}{2}~1335.7~\AA{} radiance maps. Comparing the locations of the bright knots to the \textit{SDO}/HMI LOS magnetograms, they appear to be cospatial with opposite magnetic polarity flux patches at the photospheric level. In particular, the negative flux concentrations appear to belong to the elongated, serpentine field of the EFR. 

With regard to the UV bursts, we investigated some properties of their UV emission.
The \ion{Si}{4} line profiles show a wide variety of shapes in UV bursts, and here we look for patterns that may indicate common physical mechanisms between groups of events. For this purpose, we consider relationships between the moments of the two \ion{Si}{4}~1394 and 1402~\AA{} profiles, in particular centroid, line width (full width at half maximum, FWHM), and skewness, and also relations between the moments and the line intensities. For each of the raster scans, the one hundred brightest pixels belonging to the UV bursts were considered. We defined the total intensity of each spatial pixel as the integrated intensity taken at each spectral sampling point in the spectral range $1400.8 - 1404.2$~\AA{} for the \ion{Si}{4}~1402~\AA{} line. The same pixels are investigated for the \ion{Si}{4}~1394~\AA{} line as well. For that line, the total intensity is integrated in the spectral range $1391.8 - 1395.2$~\AA{}.


Figure~\ref{fig_scatter} (top panels) presents the scatter plots of centroid vs.~skewness (left), centroid vs.~width (middle), and width vs.~skewness (right) for both the \ion{Si}{4}~1394~\AA{} (blue symbols) and 1402~\AA{} (red symbols) line profiles in the third \textit{IRIS} raster scan, at the peak of intensity (see Paper~I). (The full sets of the scatter plots for all \textit{IRIS} scans can be found in the Appendix, Figure~\ref{fig_scatter1_full}). A clear correlation is found between the centroid and the skewness (Figure~\ref{fig_scatter}, top-left panel): the more blue-shifted (red-shifted) the profile is, the larger positive (negative) skewness it has. Noticeably, most of the profiles are blue-shifted. A negative skewness for a blue-shifted profile means the line has enhanced intensity on the long wavelength side of the line, and a positive skewness for a red-shifted profile means the line has enhanced intensity on the short wavelength side of the line. Thus, the relationship in Figure~\ref{fig_scatter} implies that the burst line profiles always have enhanced emission on the side closest to zero velocity. A plausible physical explanation is that the profiles reveal plasma that is accelerated in one direction from zero velocity to some maximum velocity, with most plasma at high velocities.

There is no apparent relationship between the centroid and FWHM (Figure~\ref{fig_scatter}, top-middle panel): we only point out that the \ion{Si}{4}~1402~\AA{} line profiles have a distribution of line widths as large as that for \ion{Si}{4}~1394~\AA{} profiles during the peak of the UV bursts. As regards the relation of FWHM vs.~skewness (Figure~\ref{fig_scatter}, top-right panel), we observe that the larger line width the profile has, the larger its skewness is, especially for the \ion{Si}{4}~1394~\AA{} profiles.

\begin{figure}[b]
	\centering
	\includegraphics[scale=0.435, clip, trim=125 80 185 80]{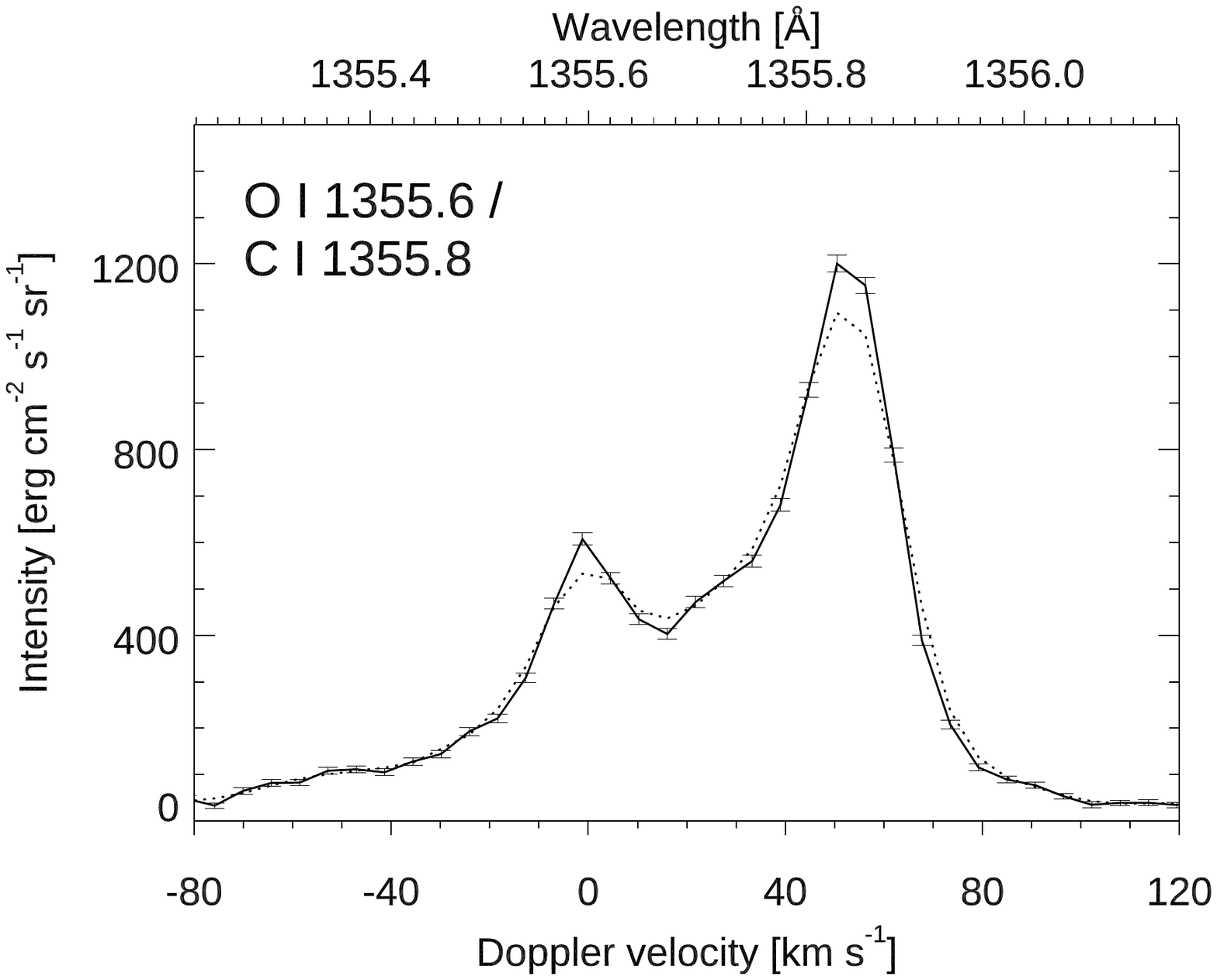}	
	\caption{Graph showing the spectrum (solid line) in the spectral region around the \ion{O}{1} 1355.60~\AA{} line, also including the \ion{C}{1} 1355.84~\AA{} line. The spectrum has been calculated as the average between three adjacent slit positions along the \textit{y} direction, centered on the pixel indicated in blue color in Paper~I, Figures~10 and~11 (UV burst core). The dotted line represents a smoothing of the same spectrum. Vertical segments represent the error bars. \label{fig_irisoici}}
\end{figure}

Similarly, we analyzed the relationship between the integrated intensity of each profile and the skewness, centroid, and FWHM, as illustrated in Figure~\ref{fig_scatter} (bottom panels) for both the \ion{Si}{4}~1394~\AA{} (blue symbols) and  1402~\AA{} (red symbols) line profiles in the UV bursts. In order to facilitate the comparison between the two lines, we divided the integrated intensity of the \ion{Si}{4}~1394~\AA{} line profiles by a factor of two while making the plots. This factor accounts for the intensity ratio between these two \ion{Si}{4} lines under the assumption that plasma is optically thin. (The full sets of the scatter plots for all IRIS scans can be found in the Appendix, Figure~\ref{fig_scatter2_full}). The scatter plots of intensity vs.~skewness (Figure~\ref{fig_scatter}, bottom-left panel) do not show any obvious trend. The profiles tend to have a larger amount of positive skewness, being irrespective of the intensity. There is no  correlation between intensity and centroid (Figure~\ref{fig_scatter}, bottom-middle panel): a few bright points are strongly red-shifted, whereas for the most part the line profiles are blue-shifted at the peak of the UV bursts, regardless of their intensity. Conversely, there is a correlation between intensity and FWHM (Figure~\ref{fig_scatter}, bottom-right panel), as larger intensities correspond to wider line profiles, although the spread is rather large.

In connection with the properties of the UV burst emission, thanks to the long exposure time we were able to study the radiation emitted in some faint lines around the \ion{O}{1} 1355.6~\AA{} line. This spectral range comprises coronal lines from hot ions such as \ion{Fe}{12} 1349.4~\AA{} and \ion{Fe}{21} 1354~\AA{}, as already mentioned. Here we resume the analysis of these spectra, whose results were briefly touched on in Paper~I. We aim at providing new information about the emission in the UV burst core during the peak (third scan), in the spatial region around the pixel indicated in blue color in Paper~I (see Figures~10 and~11), taken as the reference pixel.

\begin{figure}[b]
	\centering
	\includegraphics[scale=0.435, clip, trim=125 80 185 140]{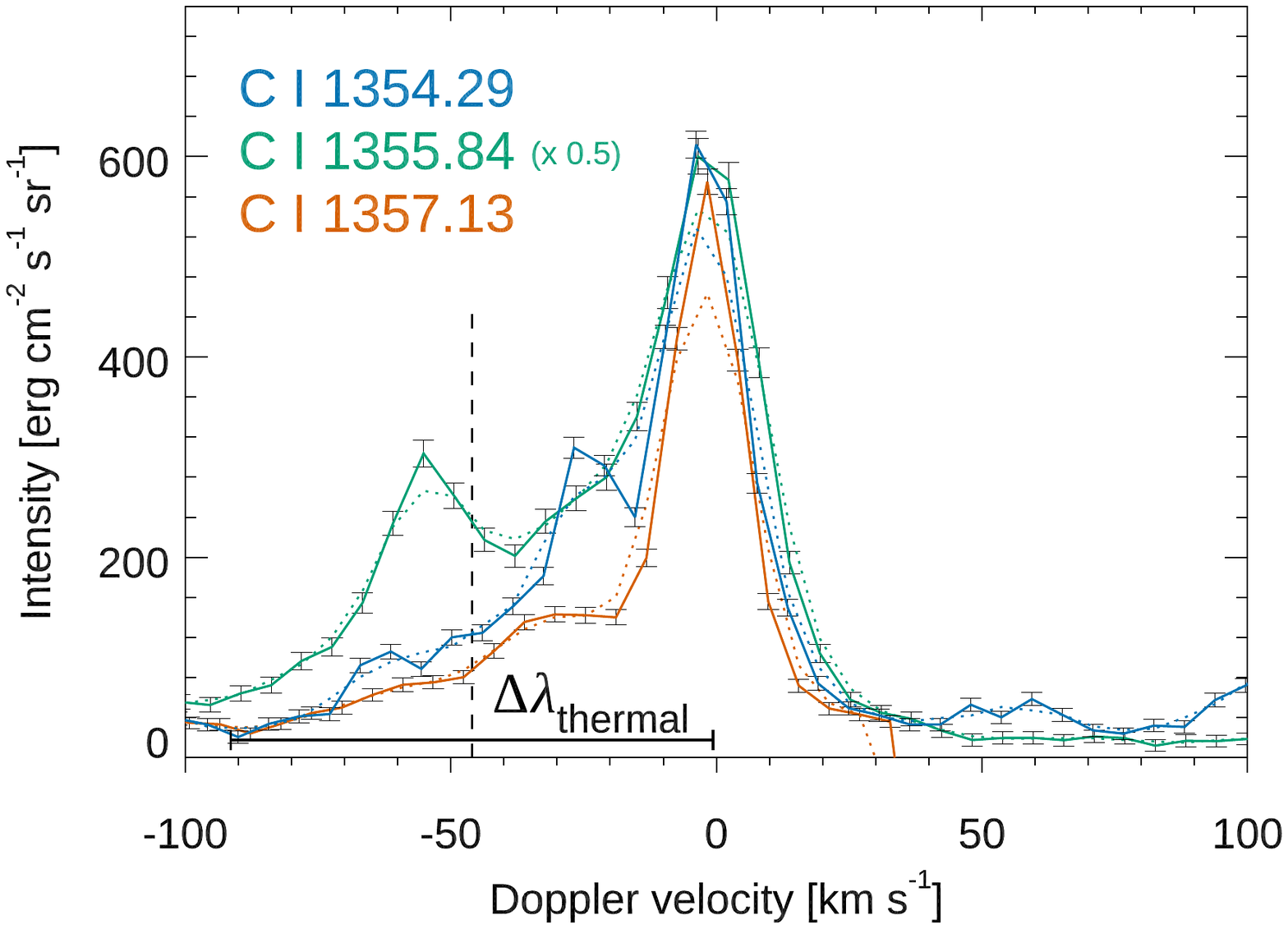}
	\caption{Graph showing the overplotted spectra (solid lines) of three \ion{C}{1} lines: 1354.29~\AA{}, 1355.84~\AA{} (divided by a factor of two), and 1357.13~\AA{}, according to the color scheme in the legend. The dotted lines represent a smoothing of the same spectra. Vertical segments represent the error bars. The spectra have been calculated as a spatial average as in Figure~\ref{fig_irisoici}. The black dashed vertical line indicates the nominal wavelength position of the \ion{Fe}{21} 1354.08~\AA{} line. The thermal width ($\Delta \lambda$) of this coronal line is represented by an horizontal segment. \label{fig_irisfexxi}}
\end{figure}

In the first place, we examined the \ion{O}{1} 1355.60~\AA{} line and its nearby \ion{C}{1} 1355.84~\AA{} line (Figure~\ref{fig_irisoici}). The spectrum was calculated as the average between three adjacent slit positions along the \textit{y} direction, centered on the reference pixel. Interestingly, in this spectrum the \ion{C}{1} intensity is larger than in the \ion{O}{1} line, with the ratio \ion{C}{1} / \ion{O}{1} being about 2. Such a behavior is not a characteristic of the reference pixel only: in the whole $3 \times 5$ pixels region around it, the ratio between the intensities of the two lines varies from 1.1 up to 2.2.

\begin{figure}[b]
	\centering
	\includegraphics[scale=0.435, clip, trim=125 80 185 80]{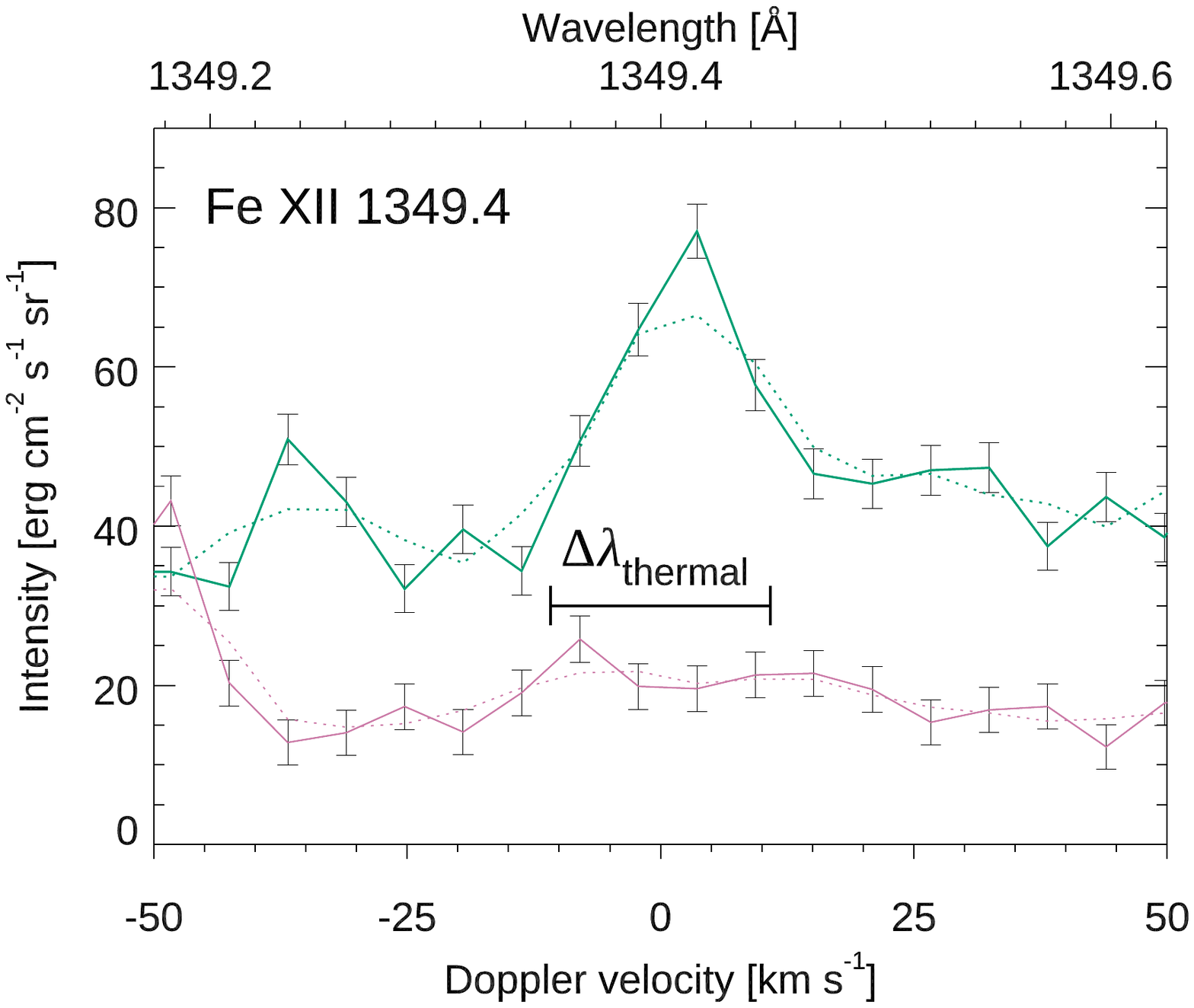}	
	\caption{Graph showing spectra (solid lines) in the spectral region around the \ion{Fe}{12} 1349.40~\AA{} line. The thermal width ($\Delta \lambda$) of this coronal line is represented by an horizontal segment. The spectrum relevant to the UV burst core (green line) has been calculated by averaging five adjacent slit positions along the \textit{y} direction and three adjacent spatial positions along the \textit{x} direction, centered at the pixel indicated in blue color in Paper~I, Figures~10 and~11 (UV burst core). For comparison, we also show a similarly averaged intensity in a quiet-Sun region (magenta line), centered in a \textit{y} position along the same slit as the center of the UV burst core. The dotted lines represent a smoothing of the same spectra. Vertical segments represent the error bars. \label{fig_irisfexii}}
\end{figure}

Then, we addressed the problem of determining whether emission in coronal lines does occur in the UV burst. To this intent, we first considered the coronal \ion{Fe}{21} 1354.08~\AA{} line (temperature formation $\log T \left[ \mathrm{K} \right] \approx 7$), which is typically observed only during flares and has a blend with the \ion{C}{1} 1354.29~\AA{} line. Figure~\ref{fig_irisfexxi} is a plot of the intensity measured in this spectral range (blue color). The position of the nominal wavelength of the \ion{Fe}{21} 1354.08~\AA{} line is indicated by a vertical dashed line, together with its large thermal FWHM of about 0.41~\AA{}. Actually, the spectrum exhibits a bump in the blue wing of the \ion{C}{1} 1354.29~\AA{} line, which could be a feature due to \ion{Fe}{21} emission. To get rid of any doubt on the nature of this bump, we compared the emission in the \ion{C}{1} 1354.29~\AA{} line with two other \ion{C}{1} lines that are comprised in the \ion{O}{1} 1355.6~\AA{} spectral range. In this regard, Figure~\ref{fig_irisfexxi} also presents the overplotted spectra for these \ion{C}{1} lines: 1355.84~\AA{} (green, divided by a factor of two) and 1357.13~\AA{} (orange). These plots clarify that the bump in the blue wing of the \ion{C}{1} 1354.29~\AA{} line is rather consistent with only \ion{C}{1} emission with extended blue wings, which is a typical feature observed in all these \ion{C}{1} lines. Thus, \ion{Fe}{21} emission appears to be absent in the UV burst.

Conversely, genuine emission in the very faint \ion{Fe}{12} 1349.4~\AA{} line (peak temperature formation $\log T \left[ \mathrm{K} \right]  \sim 6.2$, \citealp{Dere:09}) is observed in the UV burst. 
A site with enhanced emission cospatial to the UV burst seen in the other UV lines can be also found in the radiance maps for the \ion{Fe}{12} 1349~\AA{} line, although a significant contribution to the brightening in this line might be caused by the continuum enhancement in the UV burst. To verify the reliability of this signal, we analyzed the spectrum in the UV burst core, as shown in Figure~\ref{fig_irisfexii}. In order to increase the signal-to-noise ratio for this faint line, the spectrum was calculated by averaging three spatial position along the \textit{x} direction and five adjacent slit positions along the \textit{y} direction around the reference pixel. Figure~\ref{fig_irisfexii} unequivocally reveals that the line is detected in the UV burst (green profile), with a peak about a factor two higher than the background level (magenta profile). The line width appears to be slightly larger than the pure thermal FWHM (0.1~\AA{}). Therefore, even though the enhanced continuum contributes for a significant part to the brightening seen in the radiance maps, signal in the \ion{Fe}{12} 1349.4~\AA{} line is clearly detected.

\section{Discussion}

Using observations acquired by the \textit{IRIS} and \textit{SDO} satellites, we have analyzed the characteristics of UV radiation emitted during a magnetic reconnection event, occurring between a newly forming EFR and pre-existing fields in the plage of AR NOAA~12529. 

As already noticed in Paper~I, strong brightenings in the EFR area are observed in the upper atmospheric layers during the flux emergence event throughout the \textit{IRIS} observing sequence. The emission at the chromospheric level is smoothly enhanced; by contrast, repeated bursts are observed in the TR and corona. This different behavior is clearly seen in the lightcurves displayed in Figure~\ref{fig_lightcurve}: \textit{IRIS} 2796~\AA{} (chromosphere), \textit{IRIS} 1400~\AA{} (TR), and SDO/AIA 193~\AA{} (coronal level). Such a difference can be ascribed to intermittent reconnection \citep[e.g.][]{Luc:17}. 

By considering the UV emission at several wavelengths, we have been able to reconstruct the evolution of the EFR at different layers using radiance maps. The optically thin view provided by the \ion{O}{1} 1355.6~\AA{} line \citep{Lin:15} images the AFS that formed above the EFR. AFSs are typically observed in absorption in the chromospheric layers during flux emergence and reflect the serpentine nature of the emerging fields \citep{Bruzek:80,Spadaro:04,Zuccarello:05,Murabito:17}. They also can reconnect with the ambient field \citep[e.g.][]{Zuccarello:08,Tarr:14,Su:18}. Here, the AFS appears in emission when observed in the \ion{O}{1} 1355.6~\AA{} line, being brighter than the background. Some brightness enhancements are found along the AFS, suggesting the occurrence of small-scale energy release events related to the reconnection of the emerging field lines with the ambient field \citep[see, e.g.,][]{Huang:18}. The view of optically thick \ion{Mg}{2}~k and \ion{C}{2} 1335~\AA{} lines \citep{Leenarts:13,Rathore:15} offers the possibility to see the counterpart of the AFS in the upper chromosphere. Threads observed in absorption, departing from the AFS, cover the whole structure. It should be stressed that the reconstructed radiance maps image the EFR along the entire \textit{IRIS} scan: this means that the individual threads at every single slit position along the \textit{x} direction were only transiently observed. From the comparison with \textit{IRIS} SJIs in the 2796~\AA{} passband, it can be deduced that threads correspond to the surges, being viewed as individual ejections in slice imaging. The westernmost part of the UV burst, being not covered by threads, is seen as a compact brightening in the \ion{Mg}{2}~k and \ion{C}{2} 1335~\AA{} radiance maps, in the latter having a higher contrast with respect to the background. Conversely, the optically thin emission in the \ion{Si}{4}~1402 and 1394~\AA{} lines clearly illustrates the UV burst and other bright knots in the EFR area. Apparently, threads have no counterpart at \ion{Si}{4}~1402 and 1394~\AA{} formation heights, albeit fluffy elongated structures appear in the northern part of the UV burst. Finally, dark elongated absorption structures are distinctly visible in the \textit{SDO}/AIA 193 filtergrams, which are highly reminiscent of the threads seen by \textit{IRIS}. To their West, a compact brightening cospatial to the UV burst is also detected, being not obscured by such structures.

The different visibilities of the UV burst between the optically thick \ion{Mg}{2}~k and \ion{C}{2} 1335~\AA{} lines and the optically thin \ion{Si}{4}~1402 and 1394~\AA{} lines provide a strong indication that the heated plasma, which forms the brightening `hot pocket' \citep{Peter:14}, is located at atmospheric heights below the threads. In fact, numerical simulations by \citet{Nobrega:17,Nobrega:18} indicate that cool, dense surges are formed during magnetic flux emergence together with strong UV brightenings, recognized as UV bursts. The surges ejected simultaneously to the occurrence of UV bursts may obscure the brightness enhancements formed at the inner and outer footpoints of the surges, owing to the combined effect of the EFR topology and of the alignment of the LOS with respect to the EFR itself \citep{Nobrega:18}. This only affects the optically thick emission, which is blocked by the cool material of the surges, whereas the optically thin emission of \ion{Si}{4} lines can easily escape. This finding reinforces the suggestion proposed in Paper~I about a low chromospheric location of the reconnection site, which was supported by the detection of \ion{Mg}{2} triplet emission in the UV burst \citep{Pereira:15}.

In addition, we highlight that \ion{Si}{4} brightenings, represented as intensity isocontours overplotted on the \textit{SDO}/HMI magnetograms (Figure~\ref{fig_synoptic}, seventh column), are cospatial to the contact regions between the emerging, negative polarity of the EFR and the positive field of the plage. That is not restricted to the area of the UV burst, close to the pre-existing pore P$^{+}$ as described in Paper~I, but it also involves bright knots spread over the whole area where flux emergence of serpentine fields takes place. Magnetic reconnection may occur between all of the negative emerging patches of the EFR getting in contact with the positive ambient field and, hence, provides a convincing explanation for the extended intensity enhancements seen in the \ion{Si}{4} lines.

In connection with the analysis of threads, the radiance maps obtained in the wings of the \ion{Mg}{2}~k and \ion{Si}{4} 1402~\AA{} lines allow us to study their dynamics and to better understand their nature. The comparison between our observations and the numerical experiments of \citet{Nobrega:16,Nobrega:17,Nobrega:18} may help us again. Indeed, many threads appearing in the blue wing of the \ion{Mg}{2}~k line have a counterpart in the red wing, shifted to the West. Taking into account the LOS effects and the time delay between consecutive exposures, the blue/red wings counterparts can be explained as a result of the material ejected, previously exhibiting a blueshift, that later falls back down, thus being red-shifted. Furthermore, this interpretation appears to be consistent with the south-east--to--north-west orientation of the threads seen in the blue wing and the north-east--to--south-west orientation of those observed in the red wing. Such an asymmetry can be due to the bias induced by the scanning motion of the slit, over-imposed to the different kinematic state of the threads during their ascent and subsequent fall. The different lifetimes of threads between the line wings are also in agreement with this scenario. Moreover, the swaying motion of the surges found by \citet{Nobrega:16} can represent a further source of displacement along time, leading to detectable wavelength shifts. 

Interestingly, threads are observed also in the \ion{Si}{4} 1402~\AA{} line wings, although they are not evident in the line core. In this context, \citet{Nobrega:18} showed that including non-equilibrium ionization for Si and O lines in radiative MHD flux emergence numerical models strongly affects the envelope of the emerging domain, in particular of the surges. As a consequence, in the TR the corresponding emissivity in the boundaries of the surges is higher than in the region outside the emergence site. Doppler shifts due to plasma velocity during the rise and fall of the surges appear in the synthetic spectra. The visibility of the surge emission depends on the LOS and on the impact of the larger brightness of the reconnection site and of other bright features that can make negligible the contribution of the emission of the surge envelope in the total intensity integration. In addition, \citet{Nobrega:18} found that in the surge boundaries emissivity peaks at temperature values lower than the \ion{Si}{4} or \ion{O}{4} temperature formation in statistical equilibrium ($\log T \left[ \mathrm{K} \right] \sim 5$), suggesting a relationship with the rapid cooling of the plasma determined by optically thin losses and thermal conduction. In confirmation of the above interpretation, according to which \ion{Si}{4} threads are surges with enhanced TR, we carried out a preliminary analysis of the \textit{IRIS} data that shows that threads are seen in the \ion{O}{4} as well. We defer a detailed investigation of this point to a future work.  

Focusing on the properties of the UV emission in the brightening site, we performed a statistical analysis of the moments of the \ion{Si}{4} line profiles observed in the bursty region. We noted that there are no significant relationships between the intensity of the line profiles and their centroid, skewness, and line width. Instead, we found that, at the UV burst peak, the profiles are generally blue-shifted and present a strong correlation between their centroid and skewness: the more blue-shifted/red-shifted the profile is, the larger positive/negative skewness it has. Such a correlation seems to hold throughout the UV burst observations (see the Appendix). A similar trend was found in numerical simulations by \citet{Rathore:15} for the optically thick \ion{C}{2} 1335~\AA{} lines. Studying double-peak profiles, they observed an asymmetry with one peak brighter than the other. They showed that this asymmetry is correlated with a velocity gradient between the heights of formation of the peaks and the line core. A blue peak that is stronger/weaker than the red peak is correlated with a downflow/upflow of matter (relative to the velocity at the peak formation height) above the peak formation height. This result is similar to that obtained by \citet{Leenarts:13} for the optically thick \ion{Mg}{2}~h and~k lines. However, our findings are relevant to the \ion{Si}{4} line profiles, which are optically thin and usually exhibit a single peak unless different atmospheric components coexist in the same pixel. Thus, while a velocity gradient may explain the correlation between the centroid and skewness of the profiles, the physical quantities associated with such a velocity difference should be investigated.

Finally, we concentrated our attention on the UV emission at several wavelengths in the \ion{O}{1} 1355.6~\AA{} line spectral window, which is rarely observed during transient phenomena of low energy content. The ratio between the allowed \ion{C}{1}~1355.8 and intersystem \ion{O}{1} 1355.6~\AA{} lines exhibits unexpected values in the UV burst core. \ion{C}{1} / \ion{O}{1} line ratio values are always larger than unity, with peaks beyond two. This intriguing behavior was briefly discussed in Paper~I, recalling that \citet{Cheng:80} reported that \ion{C}{1}/\ion{O}{1} line ratio is remarkably enhanced during flares with respect to the typical values observed in the quiet Sun ($0.5 - 1$). An electron density enhancement by a factor of $\sim 50$ of the chromospheric plasma was proposed by these investigators as being responsible for the increase of the \ion{C}{1}/\ion{O}{1} line ratio. Recently, \citet{Lin:15} and \citet{Lin:17} studied the formation of the \ion{O}{1}~1355.6 and \ion{C}{1} 1355.8~\AA{}, respectively. In particular, \citet{Lin:17} found that the \ion{C}{1}/\ion{O}{1} total line intensity ratio is correlated with the inverse of the electron density in the mid-chromosphere, where both lines form. The larger is the electron density, the smaller are the values of the \ion{C}{1}/\ion{O}{1} line ratio, although this correlation has a considerable spread. However, this relation was derived from a quiet Sun model; thus, it could not hold in atmospheres having different properties, like those expected in the reconnection sites corresponding to UV bursts, or during flares. Therefore, the behavior of \ion{C}{1}/\ion{O}{1} line ratio still waits for further studies to be understood. 

Notably, \ion{Fe}{12} 1349.4~\AA{} emission is found in the UV burst. \citet{Testa:16} demonstrated that this forbidden \ion{Fe}{12} line, although weak, can be observed with \textit{IRIS} at high spatial resolution, provided that long enough exposure times, of the order of $30$~s, and appropriate observing modes are used, i.e., including spectral and/or spatial binning and lossless compression. Bright high-density plasma regions, somehow related to heating events, are an ideal target. Usually, the residual observed non-thermal velocities for this line are modest, of the order of $15 \,\mathrm{km \,s}^{-1}$ \citep{Testa:16}. 


As a matter of fact, a signal clearly above the background and the enhanced continuum is found in the pixels relevant to the UV burst core, with small non-thermal velocity (see Figure~\ref{fig_irisfexii}), although the latter can be systematically underestimated in \textit{IRIS} measurements due to the low signal-to-noise ratio \citep{Testa:16}. Nonetheless, this result unravels that plasma in the reconnection site is being, at least, heated up to temperatures of the order of that of the line formation ($\log T \left[ \mathrm{K} \right] \approx 6.2$). 

The detection of the \ion{Fe}{12} 1349.4~\AA{} line may explain, at least in part, the brightenings seen in the \textit{SDO}/AIA 193~\AA{} filtergrams, cospatial to the UV brightenings. In fact, \citet{ODwyer:10} showed that in ARs the \textit{SDO}/AIA 193~\AA{} channel is dominated by \ion{Fe}{12} ion (192.39, 193.51, and 195.12~\AA{} lines), whereas the continuum contribution is about an order of magnitude weaker. 
Also \citet{Sykora:11} confirmed that \ion{Fe}{12} emission is generally dominant in the 193~\AA{} channel. However, this channel has significant contribution from TR lines as well. In the observed UV burst, which appears to largely occur in the chromosphere, heating of the cool and dense material by magnetic reconnection is certainly able to produce strong emission from these TR lines, in addition to \ion{Fe}{12} emission. Nonetheless, provided that the UV burst has a pure coronal counterpart, the simultaneous observations of \textit{IRIS} and \textit{SDO}/AIA suggest that the intense brightenings observed in many \textit{SDO}/AIA EUV channels, cospatial to the UV burst (see Paper~I), are probably to some extent also due to plasma heated up to coronal temperatures. 

It is worth highlighting that a dynamic low corona is able to bring iron out of equilibrium, as shown in numerical models by \citet{Olluri:13} in the case of rapid heating induced by a nano-flare. In general, the bulk of the emission in \ion{Fe}{12} lines appears to be formed at $\log T \left[ \mathrm{K} \right] \gtrsim 6$, i.e., at coronal temperatures, even when non-equilibrium effects are included in the calculations \citep[see][]{Olluri:13,Olluri:15}. However, this aspect needs to be addressed in future studies that account for the conditions present in long-duration reconnection events, like in the UV burst here analyzed. 

On the contrary, the comparison between the shape of different \ion{C}{1} line profiles illustrated in Figure~\ref{fig_irisfexxi} allows us to safely exclude that emission in the \ion{Fe}{21} 1354~\AA{} line is present in the UV burst. In fact, the small extended bump in the blue wing of the \ion{C}{1} 1354.29~\AA{} line appears to be due to a spectral feature, i.e., blue extended wings, commonly observed in \ion{C}{1} lines during this burst. That sets an upper limit to the temperature reached in the UV burst, which must be lower than $\log T \left[ \mathrm{K} \right] \lesssim 7$. 

In a broader perspective, we can now attempt to interpret our observations in a more general context. First, we have to compare the burst here analyzed with \textit{IRIS} bombs \citep{Peter:14}. Despite many similarities, already detailed in Paper~I, the main differences between our observations and \textit{IRIS} bombs are i) the absence of \ion{Ni}{2} and \ion{Fe}{2} chromospheric absorption in the \ion{Si}{4} 1394~\AA{} line wings, and ii) the coronal counterpart of brightenings in \textit{SDO}/AIA EUV observations. 

In this respect, we can contrast the observational characteristics of the present event and those of  other kinds of transient brightenings in the TR, which have been studied in the past decades. In fact, earlier observations of the TR showed the existence of explosive events (EEs, e.g., \citealp{Dere:91}), which are highly energetic, transient ($< 600$~s), small-scale ($\sim 2\arcsec$) phenomena frequently detected  throughout the quiet and active Sun. They are seen in TR spectral lines having Doppler shifts at $\approx 100 \,\mathrm{km\,s}^{-1}$ to the red and/or blue of the rest wavelength and contain high density ($10^{12} \,\mathrm{cm}^{-3}$) plasma \citep{Doschek:16}. Some EEs also exhibit faint EUV brightenings \citep{Gupta:15}. EEs are associated with flux emergence events and were identified with magnetic reconnection episodes occurring during the cancelation of photospheric magnetic flux. However, \citet{Huang:17} recently studied EEs using \textit{IRIS} data and simultaneous observations at the Swedish Solar Telescope and showed that the intensity enhancement is about a factor of 10 with respect to the background emission, which is small compared to the increase observed in UV bursts. Moreover, only a few of them have a counterpart in the H$\alpha$ wings, suggesting the presence of cospatial surges. Taking into account that prominent plasma ejections accompany the strong long-lived UV burst here analyzed, in the form of chromospheric jet-like features like those often associated to magnetic reconnection events \citep[e.g.,][]{Nelson:13,Shelton:15,Nelson:16}, our event cannot be classified into this category. 

To our knowledge, the only other small-scale reconnection event observed by \textit{IRIS} having a coronal counterpart was the one recently described by \citet{Li:18}. Coronal temperatures were inferred from a differential emission measure analysis of the \textit{SDO}AIA channels, but we note that a clear detection of the \ion{Fe}{12} 1349.4~\AA{} emission line is a more reliable indication of coronal emission. The event occurred during small-scale cancelation between opposite polarities in the moat region around a sunspot. As in our observations, no chromospheric absorption was detected in the \ion{Si}{4} 1394~\AA{} line profile. However, it was a very short-duration event ($< 150$~s) observed with short exposure time (scan cadence 9.2~s), so that many spectral features in the faint spectral window of the \ion{O}{1} 1355.6~\AA{} line cannot be investigated. Moreover, the magnetic fluxes of the canceling polarities were about $5 \times 10^{18} \,\mathrm{Mx}$, one order of magnitude lower than in our event (see Paper~I). Furthermore, the morphology of the event studied by \citet{Li:18}, visible simultaneously in different \textit{SDO}/AIA channels with the absence of time delays between them, suggests that this feature is rather likely a TR-temperature event ($\log T \left[ \mathrm{K} \right] \sim 5$), similar to other cool  features \citep[see, e.g.,][]{Winebarger:13,Tian:14}.

Therefore, we must conclude that the present event occurs somewhere in the photosphere-to-upper chromosphere regime, rather deep in the atmosphere, and the reconnection involves all the solar atmospheric layers, being observed from the photosphere through the chromosphere and TR to the corona. As mentioned by \citet{Young:18}, the particular location of the reconnection site and also the line-of-sight effects may be
responsible for differences in burst signatures such as the variations in \ion{Si}{4} profiles. Furthermore, it is worth stressing that the different magnetic topology in various reconnection environments may result in differences in the extent of the atmospheric levels involved by the eruptive phenomena occurring during reconnection and, hence, in the spectral features linked to different plasma densities.

\section{Conclusions}

In Paper~I and in the present paper, we have investigated the small-scale eruptive phenomena occurring in the solar atmosphere due to the interaction between new emerging magnetic flux and ambient fields in the plage of AR NOAA~12529. In particular, in this paper we have detailed the properties of the UV emission observed by \textit{IRIS} in the upper chromosphere and in the TR of the EFR site. 

The most notable finding is the detection of \ion{Fe}{12} 1349~\AA{} emission in the core of the brightening site. This represents a pure coronal counterpart for the event and indicates that plasma is heated up to $\log T \left[ \mathrm{K} \right] \gtrsim 6$ in the UV burst. In contrast, the absence of \ion{Fe}{21} 1354~\AA{} emission implies that temperature remains lower than $\log T \left[ \mathrm{K} \right] \lesssim 7$.

Another characteristic observed in the UV burst core is an unexpected behaviour of the relative ratio between the intensities of the \ion{C}{1}~1355.8 and \ion{O}{1} 1355.6~\AA{} lines. In the bursting region, the latter is usually a factor of two weaker than the \ion{C}{1} line, as commonly observed during solar flares \citep{Cheng:80}, and a factor $2-4$ larger than the ratio found in the quiet Sun. This has been linked to electron density increase, although a theoretical understanding is lacking.

\ion{Si}{4} emission is observed not only in the prominent compact brightening, visible throughout the \textit{IRIS} observing sequence, but also in bright knots in spatial correspondence with tiny polarity inversion lines spread over the area occupied by the serpentine fields brought into the photosphere by flux emergence. These knots are found only in the spectroheliograms, owing to the optical thinness of \ion{Si}{4} lines, but cannot be observed in the \textit{IRIS} SJ 1400~\AA{} broad-band images, which include many other contributions. 

Again concerning \ion{Si}{4} emission, there is a strong correlation between the centroid and the skewness of both \ion{Si}{4} 1394 and 1402~\AA{} lines in the profiles relevant to the UV burst: the more Doppler-shifted is the profiles, the larger its skewness is. In more detail, the more blue-shifted (red-shifted) the profile is, the larger positive (negative) skewness it has. This relationship indicates that the UV burst line profiles always have enhanced emission on the side closest to zero velocity, suggesting the presence of velocity gradients. A possible explanation might be that plasma is accelerated in one direction from zero velocity to some maximum velocity, with most plasma at high velocities.

We also observe the occurrence of repeated surge-like plasma ejections departing from the AFS above the EFR. Interestingly, surges observed in absorption in chromospheric lines have a prominent counterpart in emission in the wings of \ion{Mg}{2} and \ion{Si}{4} lines, appearing in the form of threads. They also show a blue-to-red asymmetry in the line wings, which can be explained in terms of different kinematic states of the plasma ejecta. Indeed, recent numerical simulations predict that surges can appear in emission in the TR, given that non-equilibrium ionization occurs in a sheath enveloping the surges during flux emergence and subsequent magnetic reconnection with ambient fields \citep{Nobrega:18}. This aspect deserves further investigations, including analyses in other TR lines.

Comparing the properties of this burst with those of other transient, small-scale events observed in the TR, such as \textit{IRIS} bombs and EEs, we suggest that the configuration of the magnetized environment where flux emergence occurs may have an impact on the height of the reconnection site and on the amount of energy released, thus determining the presence of hot plasma with coronal temperature during such a kind of phenomena.

Most important, evidence has been provided that long-lasting, intermittent brightenings and repeated plasma ejections observed in the EFR site result from magnetic reconnection in the low atmosphere. This highlights that such a fundamental process plays a major role in the solar atmosphere and, possibily, might contribute to coronal heating as suggested by recent observations \citep{Chitta:17,Smitha:18} and numerical simulations \citep{Ni:15,Ni:16,Alvarez:17}. A significant advance in the understanding of magnetic reconnection will be possible with future high-resolution, multi-wavelength observations with the next generation solar observatories, such as the Solar Orbiter space mission \citep{Muller:13} and the large-aperture ground-based 4m-class aperture telescopes DKIST \citep[Daniel K. Inouye Solar Telescope,][]{Keil:10} and EST \citep[European Solar Telescope,][]{Collados:10}.

\acknowledgments

The authors would like to thank the anonymous referee for his/her helpful comments. 
The research leading to these results has received funding from the European Commissions Seventh Framework Programme under the grant agreement no.~312495 (SOLARNET project) and from the European Union's Horizon 2020 research and innovation programme under the grant agreements no.~739500 (PRE-EST project) and no.~824135 (SOLARNET project). This work was also supported by the Italian MIUR-PRIN grant 2012P2HRCR on \textit{The active Sun and its effects on Space and Earth climate}, by the Universit\`{a} degli Studi di Catania (Piano per la Ricerca Universit\`{a} di Catania 2016-2018 -- Linea di intervento~1 ``Chance''; Linea di intervento~2 ``Dotazione ordinaria''), by the Istituto Nazionale di Astrofisica (PRIN INAF 2014), and by Space Weather Italian COmmunity (SWICO) Research Program. PRY acknowledges funding from NASA grant NNX15AF48G, and he thanks ISSI Bern for supporting the International Team Meeting ``Solar UV bursts -- a new insight to magnetic reconnection''. \textit{IRIS} is a NASA small explorer mission developed and operated by LMSAL with mission operations executed at NASA Ames Research center and major contributions to downlink communications funded by ESA and the Norwegian Space Centre. The \textit{SDO}/HMI and \textit{SDO}/AIA data used in this paper are courtesy of NASA/\textit{SDO} and the HMI and AIA science teams. Use of NASA's Astrophysical Data System is gratefully acknowledged.

\facility{\textit{IRIS}, \textit{SDO} (HMI, AIA)}

\appendix
\section{Scatter plots - full sets}

\begin{figure}[t]
	\centering
	\includegraphics[scale=0.725, clip, trim= 0 32 200 140]{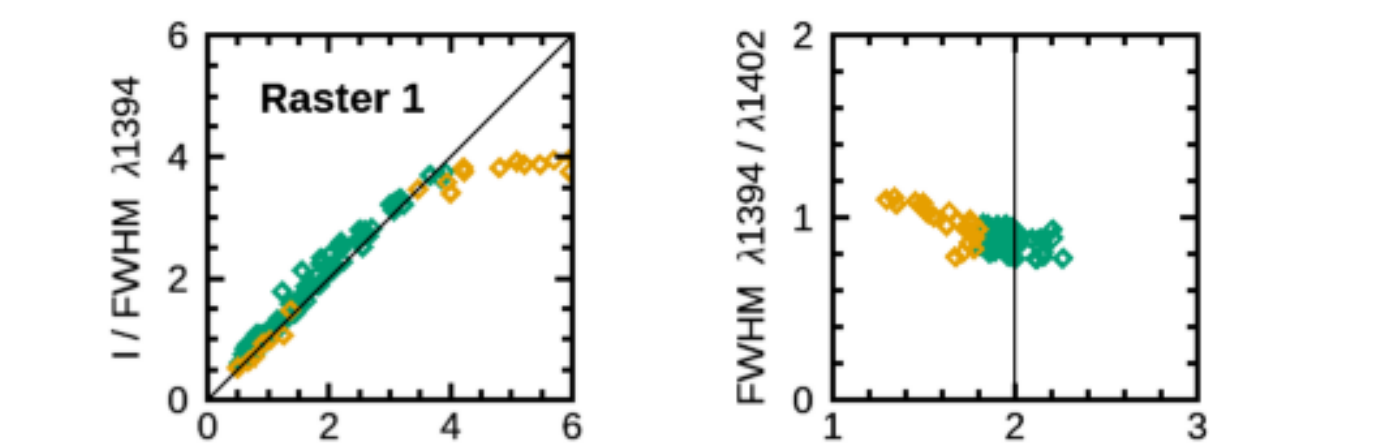}
	\includegraphics[scale=0.725, clip, trim= 0 32 200 140]{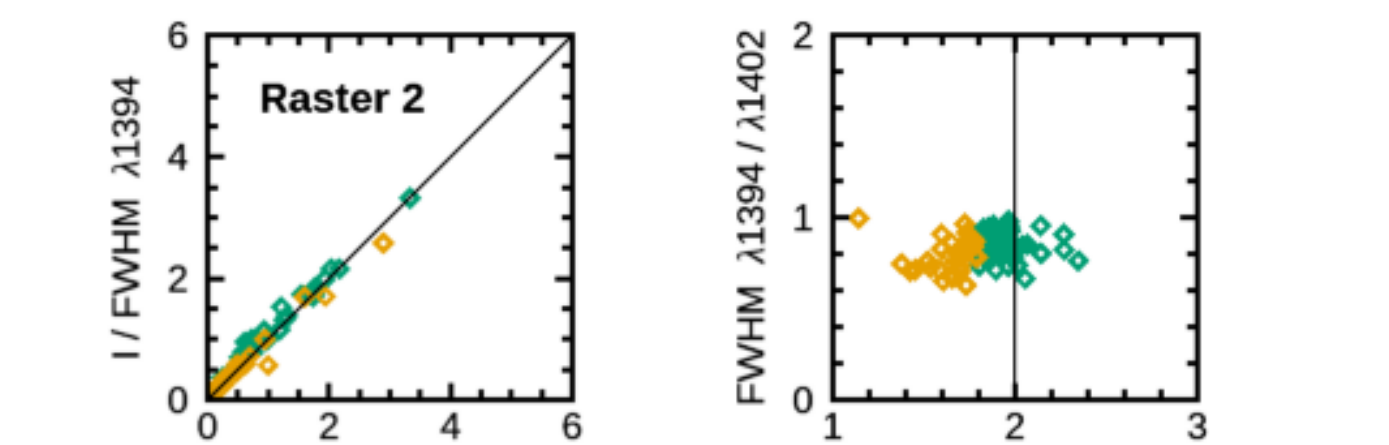}
	\includegraphics[scale=0.725, clip, trim= 0 32 200 140]{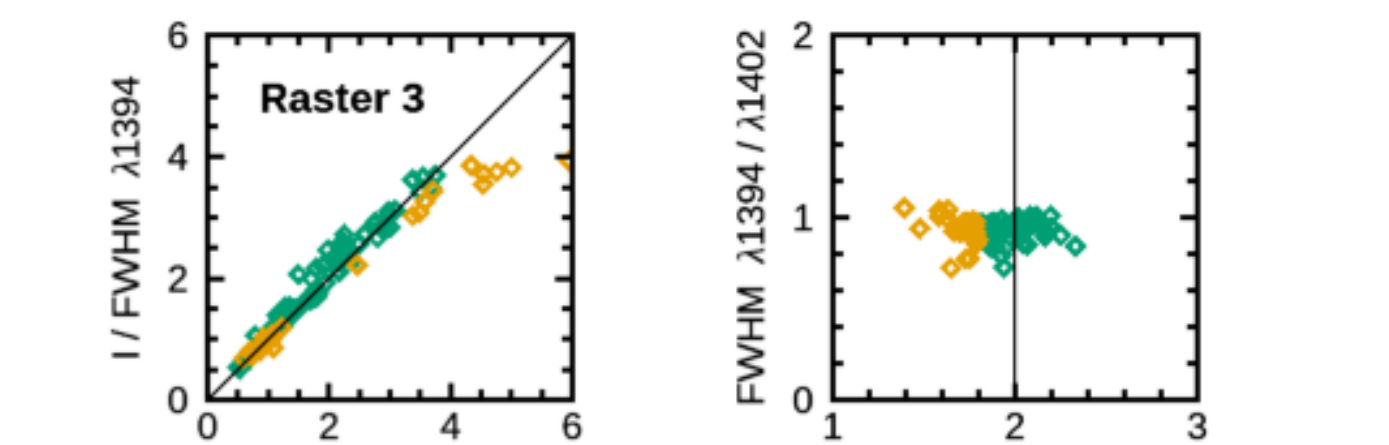}
	\includegraphics[scale=0.725, clip, trim= 0 32 200 140]{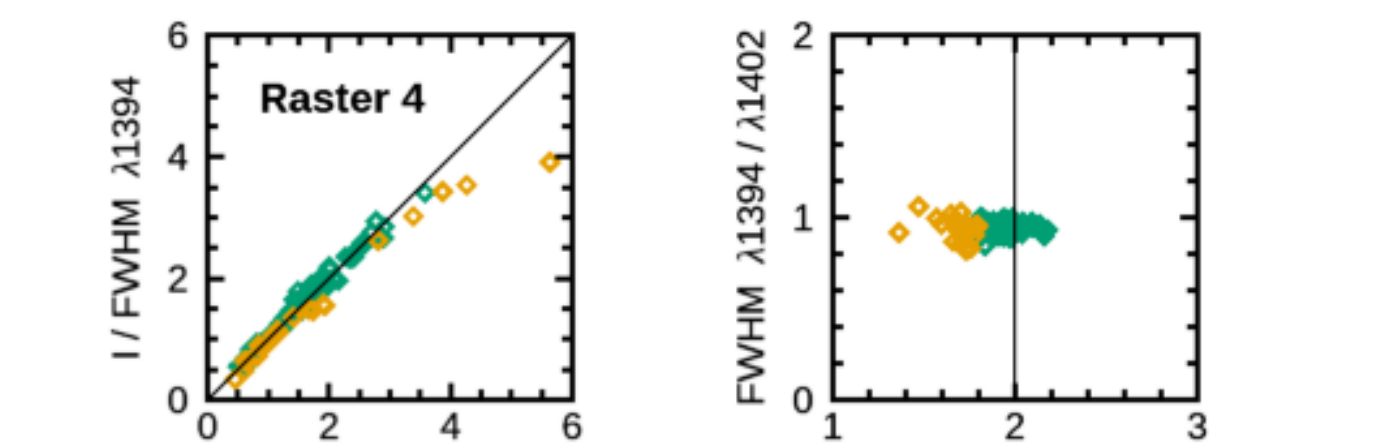}
	\includegraphics[scale=0.725, clip, trim= 0 32 200 140]{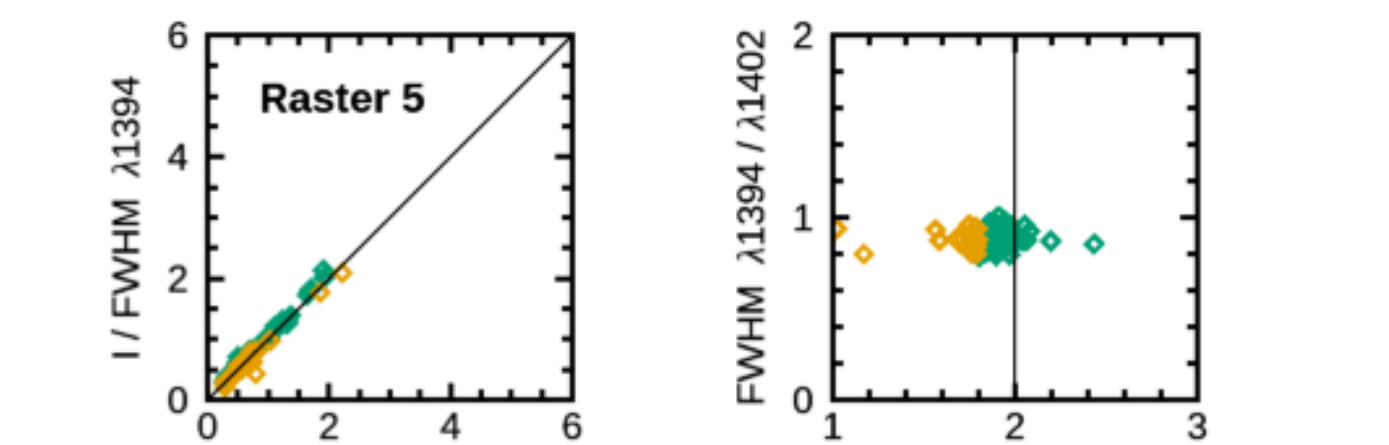}
	\includegraphics[scale=0.725, clip, trim= 0 10 200 140]{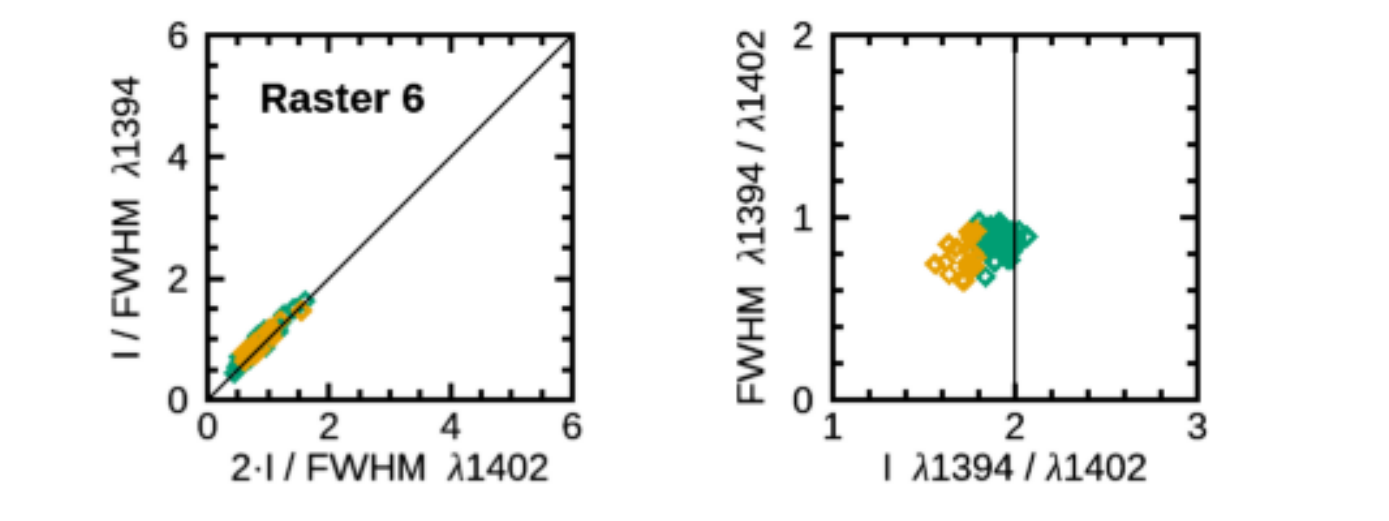}
	\caption{Scatter plots of the ratio between intensities and FWHM for the \ion{Si}{4}~1394 and 1402~\AA{} line profiles. Left column: ratio between intensity and FWHM for each individual line, \ion{Si}{4}~1402~\AA{} vs.~\ion{Si}{4}~1394~\AA{} line. Right column: ratio between FWHM for both lines vs.~ratio between intensities for both lines. Yellow color: pixels with \ion{Si}{4}~1394/1402 intensities ratio lesser than 1.8. \label{fig_scatter_thinness}}
\end{figure}

Figure~\ref{fig_scatter_thinness} is an illustration of the scatter plots of the ratio between intensities and FWHM for both the \ion{Si}{4}~1394 and 1402~\AA{} line profiles relevant to the UV burst core. In the left column we plot the ratio between intensity and FWHM for each individual line, \ion{Si}{4}~1402~\AA{} vs.~\ion{Si}{4}~1394~\AA{} line. The right column shows the ratio between FWHM for both lines vs.~ratio between intensities for both lines. Yellow symbols indicate pixels with ratio between intensities lower than 1.8. The graphs clarify that the optically thin approximation is a very good guess for the most part of the pixels: green symbols align well along $x=y$ line, corresponding to ratio equal two. However, there are two distinct groups of yellow symbols that depart from this value. The first is composed by pixels with high intensities, which can be affected by saturation in the \ion{Si}{4}~1394~\AA{} line, especially during the first \textit{IRIS} scan owing to the long exposure time. The second group consists of pixels with relatively low intensities, increasing in number when the burst is less strong. Note that pixels with ratio between intensities significantly larger than two may be affected by resonant scattering, as pointed out by \citet{Gontikakis:18}.

\begin{figure*}[t]
	\centering
	\includegraphics[scale=0.675, clip, trim= 0 30 0 140]{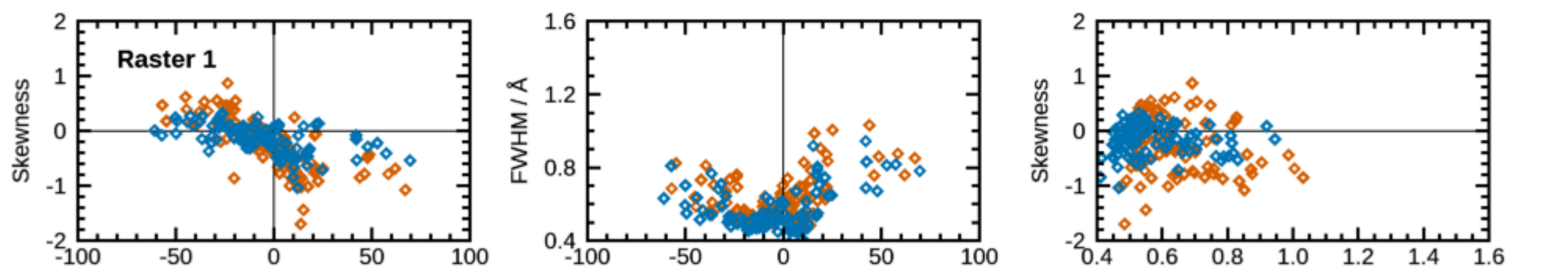}
	\includegraphics[scale=0.675, clip, trim= 0 30 0 140]{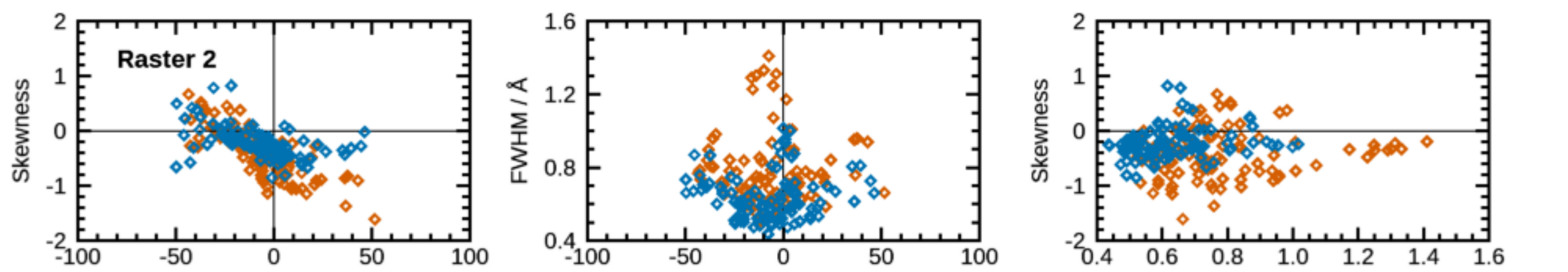}
	\includegraphics[scale=0.675, clip, trim= 0 30 0 140]{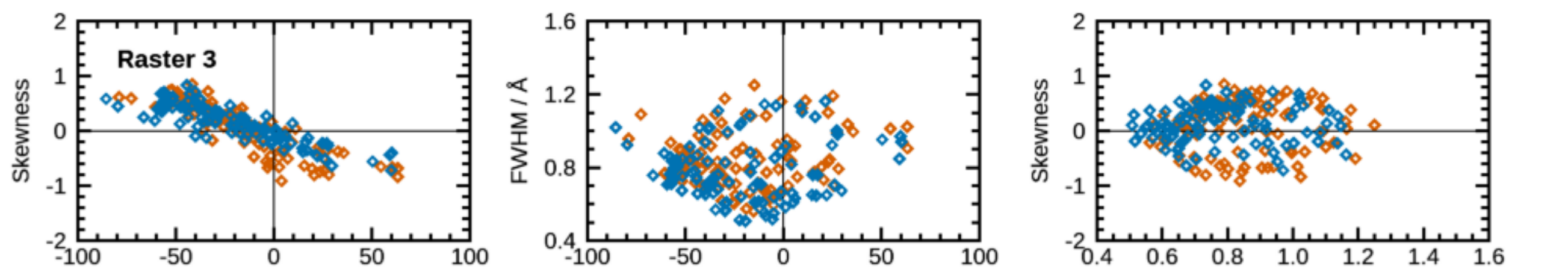}
	\includegraphics[scale=0.675, clip, trim= 0 30 0 140]{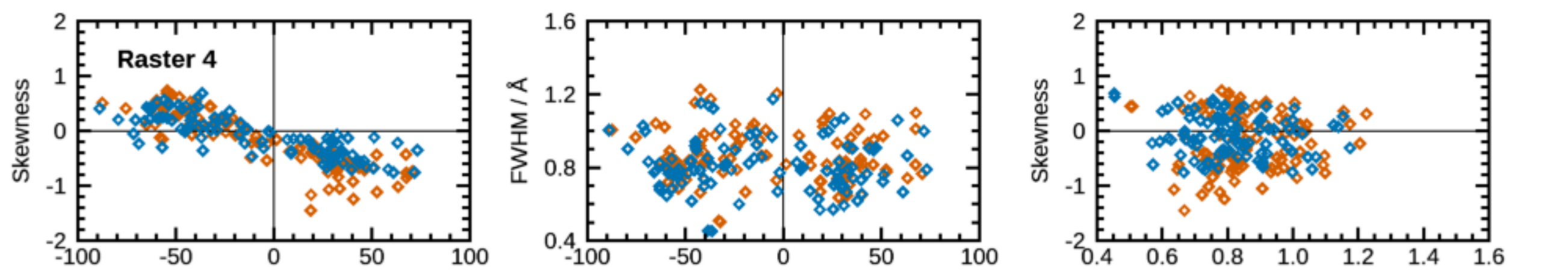}
	\includegraphics[scale=0.675, clip, trim= 0 30 0 140]{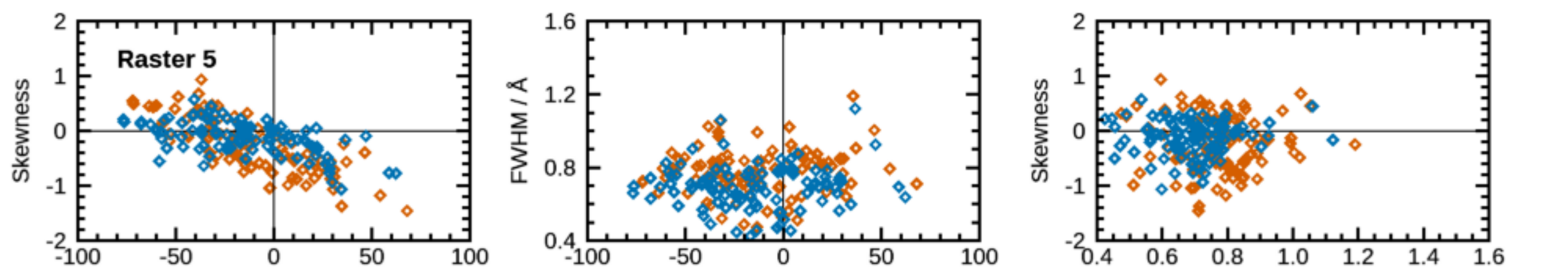}
	\includegraphics[scale=0.675, clip, trim= 0 10 0 140]{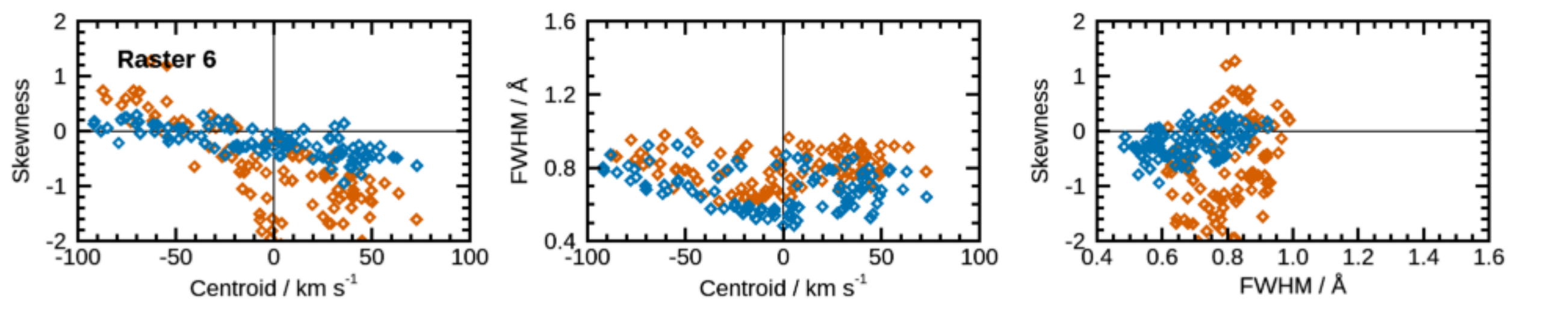}
	\caption{Plots of centroid vs.~skewness (first column), centroid vs.~FWHM (second column), and FWHM vs.~skewness (third column) for both the \ion{Si}{4}~1394~\AA{} (blue symbols) and 1402~\AA{} (red symbols) line profiles for each \textit{IRIS} raster scan. \label{fig_scatter1_full}}
\end{figure*}

In Figure~\ref{fig_scatter1_full} we analyze the correlation between centroid, skewness, and line width (FWHM). We noticed a correlation between the centroid and the skewness (the more blue- or red-shifted the profile is, the larger positive/negative skewness it has) of the profiles relevant to the UV burst at its peak. This trend is visible during throughout the \textit{IRIS} observations (see Figure~\ref{fig_scatter1_full}, left column), although only in the \ion{Si}{4}~1402~\AA{} line when the area with enhanced intensity shrinks (fifth and sixth scans). At those times, the centroid distribution is much more scattered. The \ion{Si}{4}~1402~\AA{} line profiles are generally wider than the \ion{Si}{4}~1394~\AA{} profiles (Figure~\ref{fig_scatter1_full}, middle column), except during the peak of the UV bursts (third and fourth scans), when the distribution of the FWHM is almost identical for both lines. Similarly, the observed correlation between FWHM and skewness (larger line widths correspond to a larger skewness of the profiles) holds only at the peak of the UV bursts, during the third scan (Figure~\ref{fig_scatter1_full}, right column). During the fourth scan, the distribution of skewness is almost identical for both lines, without an obvious trend. For the remaining scans the distribution of skewness is rather different between the lines. However, the \ion{Si}{4}~1402~\AA{} line profiles with larger FWHM have a larger amount of (negative) skewness, in particular in the sixth scan.

\begin{figure*}[t]
	\centering
	\includegraphics[scale=0.675, clip, trim= 0 30 0 140]{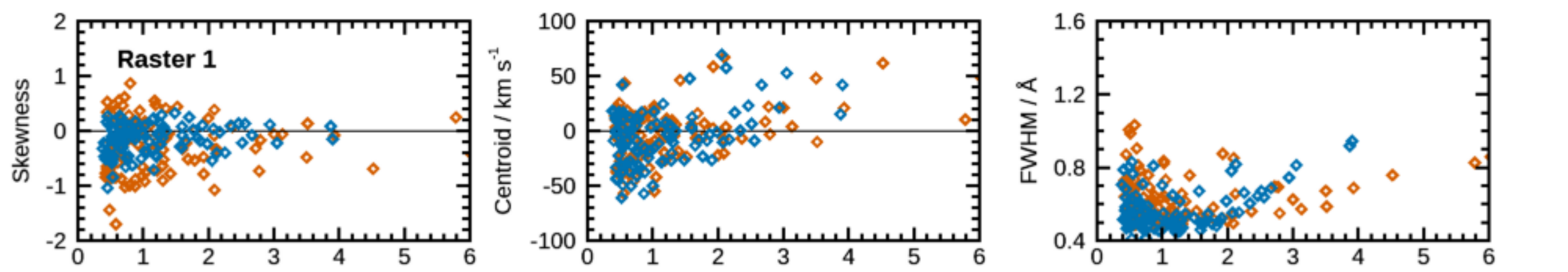}
	\includegraphics[scale=0.675, clip, trim= 0 30 0 140]{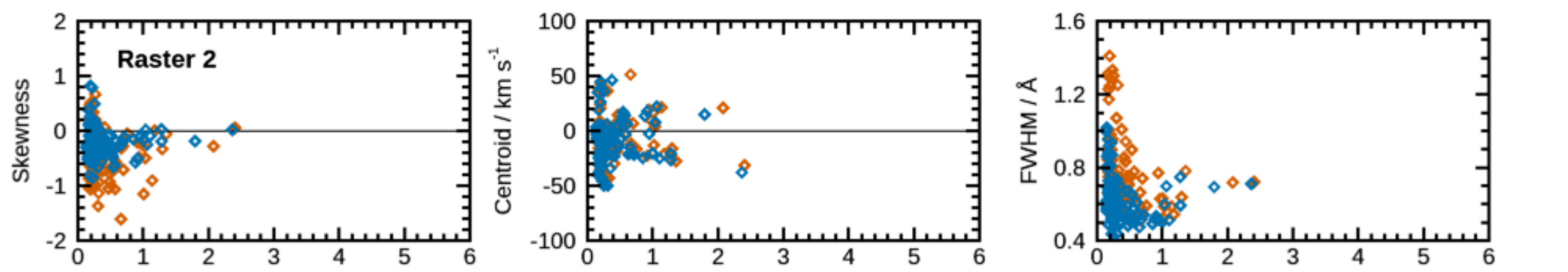}
	\includegraphics[scale=0.675, clip, trim= 0 30 0 140]{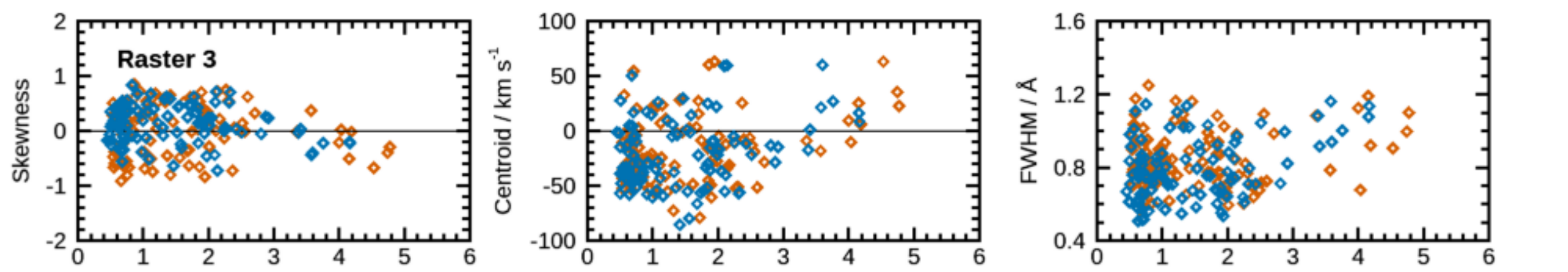}
	\includegraphics[scale=0.675, clip, trim= 0 30 0 140]{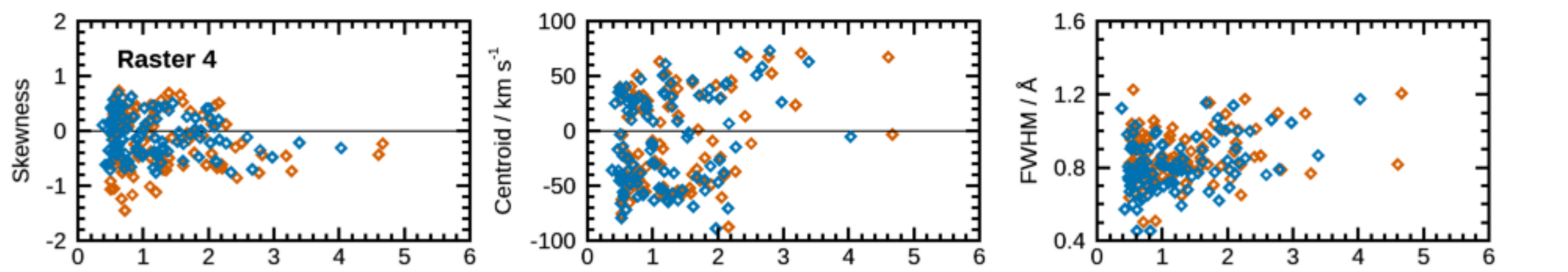}
	\includegraphics[scale=0.675, clip, trim= 0 30 0 140]{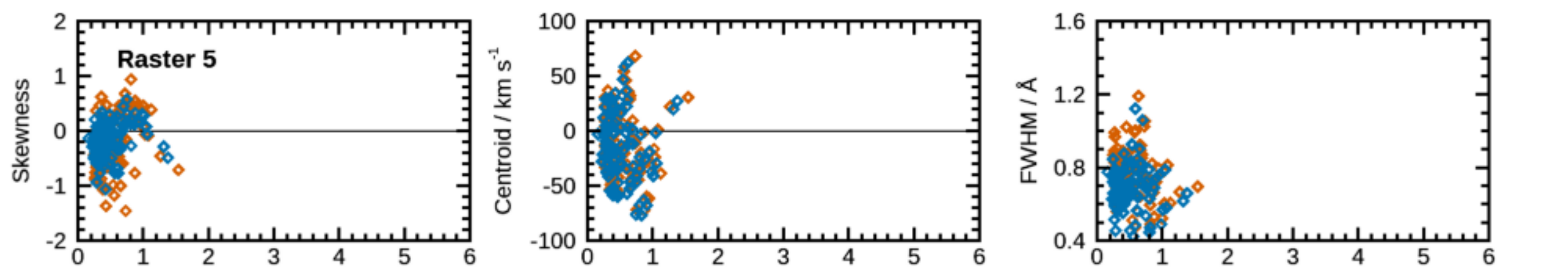}
	\includegraphics[scale=0.675, clip, trim= 0 10 0 140]{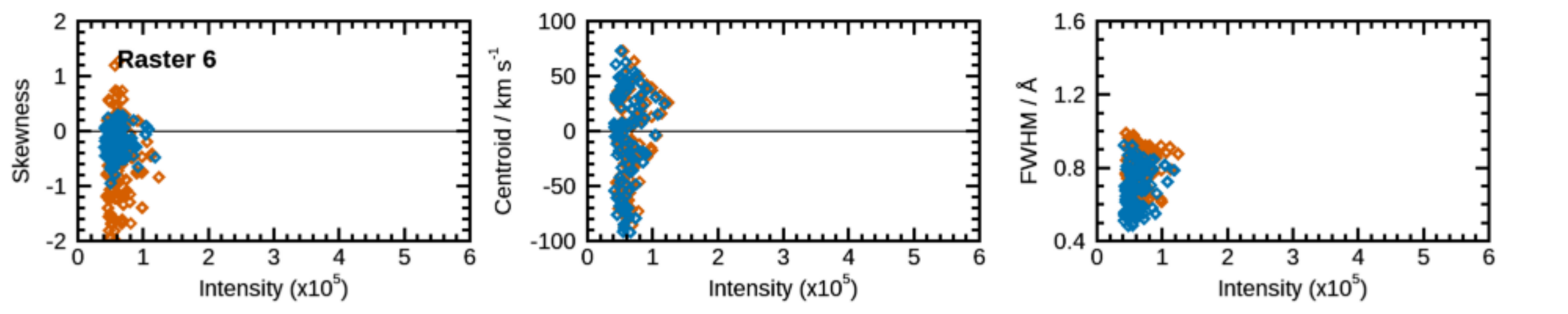}
	\caption{Plots of integrated intensity vs.~skewness (first column), vs.~centroid (second column), and vs.~FWHM (third column) for both the \ion{Si}{4}~1394~\AA{} (blue symbols) and 1402~\AA{} (red symbols) line profiles for each \textit{IRIS} raster scan. \label{fig_scatter2_full}}
\end{figure*}

As concerns the relationship between intensity and skewness, centroid, and FWHM, we note the absence of a systematic trend in the scatter plots of intensity vs.~skewness (Figure~\ref{fig_scatter2_full}, left column). They are very similar for both lines, except during the sixth scans, when the \ion{Si}{4}~1402~\AA{} profiles exhibit a much larger spread in skewness, shifted toward more negative values. At the peak of the UV bursts (third scan), the profiles tend to have a larger amount of positive skewness. The distributions of intensity vs.~centroid (Figure~\ref{fig_scatter2_full}, middle column) for both lines are akin to each other at every time. However, there appears no correlation between these two parameters. In the first, third, and fourth scans a few bright points exhibit a strong redshift. At the peak (third scan), the majority of the line profiles is blue-shifted, regardless of their intensity. In the same scans (first, third, and fourth), we find that larger intensities correspond to wider line profiles, with a large spread (Figure~\ref{fig_scatter2_full}, right column). For the rest, we again observe the general trend that \ion{Si}{4}~1402~\AA{} line profiles have a larger FWHM than \ion{Si}{4}~1394~\AA{} profiles, except during the peak of the UV bursts (third and fourth scans).

\begin{thebibliography}


\bibitem[Alvarez Laguna et al.(2017)]{Alvarez:17} Alvarez Laguna, A., Lani, A., Mansour, N.~N., Deconinck, H., \& Poedts, S.\ 2017, \apj, 842, 117








\bibitem[Bruzek(1980)]{Bruzek:80} Bruzek, A.\ 1980, Solar and Interplanetary Dynamics, 91, 203 

\bibitem[Centeno et al.(2017)]{Centeno:17} Centeno, R., Blanco Rodr{\'{\i}}guez, J., Del Toro Iniesta, J.~C., et al.\ 2017, \apjs, 229, 3


\bibitem[Cheng et al.(1980)]{Cheng:80} Cheng, C.~C., Feldman, U., \& Doschek, G.~A.\ 1980, \aap, 86, 377 

\bibitem[Cheung \& Isobe(2014)]{Cheung:14} Cheung, M.~C.~M., \& Isobe, H.\ 2014, Living Reviews in Solar Physics, 11, 3



\bibitem[Chitta et al.(2017)]{Chitta:17} Chitta, L.~P., Peter, H., Young, P.~R., \& Huang, Y.-M.\ 2017, \aap, 605, A49 

\bibitem[Collados et al.(2010)]{Collados:10} Collados, M., Bettonvil, F., Cavaller, L., \& EST Team 2010, AN, 331, 615

\bibitem[de la Cruz Rodr{\'{\i}}guez et al.(2015)]{Jaime:15} de la Cruz Rodr{\'{\i}}guez, J., Hansteen, V., Bellot-Rubio, L., \& Ortiz, A.\ 2015, \apj, 810, 145 


\bibitem[De Pontieu et al.(2014)]{DePontieu:14} De Pontieu, B., Title, A.~M., Lemen, J.~R., et al.\ 2014, \solphys, 289, 2733

\bibitem[Dere et al.(2009)]{Dere:09} Dere, K.~P., Landi, E., Young, P.~R., et al.\ 2009, \aap, 498, 915

\bibitem[Dere et al.(1991)]{Dere:91} Dere, K.~P., Bartoe, J.-D.~F., Brueckner, G.~E., Ewing, J., \& Lund, P.\ 1991, \jgr, 96, 9399 

\bibitem[Doschek et al.(2016)]{Doschek:16} Doschek, G.~A., Warren, H.~P., \& Young, P.~R.\ 2016, \apj, 832, 77 


\bibitem[Galsgaard et al.(2007)]{Galsgaard:07} Galsgaard, K., Archontis, V., Moreno-Insertis, F., \& Hood, A.~W.\ 2007, \apj, 666, 516 

\bibitem[Galsgaard et al.(2005)]{Galsgaard:05} Galsgaard, K., Moreno-Insertis, F., Archontis, V., \& Hood, A.\ 2005, \apjl, 618, L153 

\bibitem[Grubecka et al.(2016)]{Grubecka:16} Grubecka, M., Schmieder, B., Berlicki, A., et al.\ 2016, \aap, 593, A32 

\bibitem[Gontikakis \& Vial(2018)]{Gontikakis:18} Gontikakis, C., \& Vial, J.-C.\ 2018, \aap, \textit{in press}

\bibitem[Guglielmino et al.(2018a)]{Guglielmino:18} Guglielmino, S.~L., Zuccarello, F., Young, P.~R., Murabito, M. \& Romano, P.\ 2018, \apj, 856, 127

\bibitem[Guglielmino et al.(2018b)]{Guglielmino:18UF} Guglielmino, S.~L., Romano, P., Ruiz~Cobo, B., Zuccarello, F. \&  Murabito, M.\ 2018, \apj, \textit{submitted}

\bibitem[Guglielmino et al.(2017)]{Guglielmino:17} Guglielmino, S.~L., Romano, P., \& Zuccarello, F.\ 2017, \apjl, 846, L16 

\bibitem[Guglielmino(2012)]{Guglielmino:12} Guglielmino, S.~L.\ 2012, 4th Hinode Science Meeting: Unsolved Problems and Recent Insights, 455, 109

\bibitem[Guglielmino et al.(2010)]{Guglielmino:10} Guglielmino, S.~L., Bellot Rubio, L.~R., Zuccarello, F., et al.\ 2010, \apj, 724, 1083

\bibitem[Guglielmino et al.(2008)]{Guglielmino:08} Guglielmino, S.~L., Zuccarello, F., Romano, P., \& Bellot Rubio, L.~R.\ 2008, \apjl, 688, L111

\bibitem[Gupta \& Tripathi(2015)]{Gupta:15} Gupta, G.~R., \& Tripathi, D.\ 2015, \apj, 809, 82


\bibitem[Heyvaerts et al.(1977)]{Heyvaerts:77} Heyvaerts, J., Priest, E.~R., \& Rust, D.~M.\ 1977, \apj, 216, 123


\bibitem[Hong et al.(2017)]{Hong:17} Hong, J., Ding, M.~D., \& Cao, W.\ 2017, \apj, 838, 101 



\bibitem[Huang et al.(2018)]{Huang:18} Huang, Z., Mou, C., Fu, H., et al.\ 2018, \apjl, 853, L26

\bibitem[Huang et al.(2017)]{Huang:17} Huang, Z., Madjarska, M.~S., Scullion, E.~M., et al.\ 2017, \mnras, 464, 1753






\bibitem[Jiang et al.(2015)]{Jiang:15} Jiang, F., Zhang, J., \& Yang, S.\ 2015, \pasj, 67, 78

\bibitem[Keil et al.(2010)]{Keil:10} Keil, S. L., Rimmele, T. R., Wagner, J., \& ATST Team 2010, AN, 331, 609

\bibitem[Kim et al.(2015)]{Kim:15} Kim, Y.-H., Yurchyshyn, V., Bong, S.-C., et al.\ 2015, \apj, 810, 38 


\bibitem[Leenaarts et al.(2013)]{Leenarts:13} Leenaarts, J., Pereira, T.~M.~D., Carlsson, M., Uitenbroek, H., \& De Pontieu, B.\ 2013, \apj, 772, 90 

\bibitem[Lemen et al.(2012)]{Lemen:12} Lemen, J. R., Title, A. M., Akin, D. J., et al. 2012, Sol. Phys. 275, 17

\bibitem[Li et al.(2018)]{Li:18} Li, D., Li, L., \& Ning, Z.\ 2018, \mnras, 479, 2382 

\bibitem[Libbrecht et al.(2017)]{Libbrecht:17} Libbrecht, T., Joshi, J., de~la~Cruz~Rodr{\'{\i}}guez, J., Leenaarts, J., \& Ramos, A.~A.\ 2017, \aap, 598, A33 

\bibitem[Lin et al.(2017)]{Lin:17} Lin, H.-H., Carlsson, M., \& Leenaarts, J.\ 2017, \apj, 846, 40 

\bibitem[Lin \& Carlsson(2015)]{Lin:15} Lin, H.-H., \& Carlsson, M.\ 2015, \apj, 813, 34



\bibitem[MacTaggart et al.(2015)]{David:15} MacTaggart, D., Guglielmino, S.~L., Haynes, A.~L., Simitev, R., \& Zuccarello, F.\ 2015, \aap, 576, A4 


\bibitem[Mart{\'{\i}}nez-Sykora et al.(2011)]{Sykora:11} Mart{\'{\i}}nez-Sykora, J., De Pontieu, B., Testa, P., \& Hansteen, V.\ 2011, \apj, 743, 23 




\bibitem[Moreno-Insertis et al.(2008)]{Fernando:08} Moreno-Insertis, F., Galsgaard, K., \& Ugarte-Urra, I.\ 2008, \apjl, 673, L211 

\bibitem[Murabito et al.(2017)]{Murabito:17} Murabito, M., Romano, P., Guglielmino, S.~L., \& Zuccarello, F.\ 2017, \apj, 834, 76

\bibitem[M{\"u}ller et al.(2013)]{Muller:13} M{\"u}ller, D., Marsden, R.~G., St.~Cyr, O.~C., \& Gilbert, H.~R.\ 2013, \solphys, 285, 25

\bibitem[Nelson et al.(2017)]{Nelson:17} Nelson, C.~J., Freij, N., Reid, A., et al.\ 2017, \apj, 845, 16 

\bibitem[Nelson et al.(2016)]{Nelson:16} Nelson, C.~J., Doyle, J.~G., \& Erd{\'e}lyi, R.\ 2016, \mnras, 463, 2190 

\bibitem[Nelson \& Doyle(2013)]{Nelson:13} Nelson, C.~J., \& Doyle, J.~G.\ 2013, \aap, 560, A31 

\bibitem[Ni et al.(2016)]{Ni:16} Ni, L., Lin, J., Roussev, I.~I., \& Schmieder, B.\ 2016, \apj, 832, 195 

\bibitem[Ni et al.(2015)]{Ni:15} Ni, L., Kliem, B., Lin, J., \& Wu, N.\ 2015, \apj, 799, 79 

\bibitem[N{\'o}brega-Siverio et al.(2018)]{Nobrega:18} N{\'o}brega-Siverio, D., Moreno-Insertis, F., \& Mart{\'{\i}}nez-Sykora, J.\ 2018, \apj, 858, 8 

\bibitem[N{\'o}brega-Siverio et al.(2017)]{Nobrega:17} N{\'o}brega-Siverio, D., Mart{\'{\i}}nez-Sykora, J., Moreno-Insertis, F., \& Rouppe van der Voort, L.\ 2017, \apj, 850, 153 

\bibitem[N{\'o}brega-Siverio et al.(2016)]{Nobrega:16} N{\'o}brega-Siverio, D., Moreno-Insertis, F., \& Mart{\'{\i}}nez-Sykora, J.\ 2016, \apj, 822, 18 

\bibitem[O'Dwyer et al.(2010)]{ODwyer:10} O'Dwyer, B., Del Zanna, G., Mason, H.~E., Weber, M.~A., \& Tripathi, D.\ 2010, \aap, 521, A21 

\bibitem[Olluri et al.(2015)]{Olluri:15} Olluri, K., Gudiksen, B.~V., Hansteen, V.~H., \& De Pontieu, B.\ 2015, \apj, 802, 5 

\bibitem[Olluri et al.(2013)]{Olluri:13} Olluri, K., Gudiksen, B.~V., \& Hansteen, V.~H.\ 2013, \aj, 145, 72 

\bibitem[Ortiz et al.(2016)]{Ortiz:16} Ortiz, A., Hansteen, V.~H., Bellot Rubio, L.~R., et al.\ 2016, \apj, 825, 93

\bibitem[Ortiz et al.(2014)]{Ortiz:14} Ortiz, A., Bellot Rubio, L.~R., Hansteen, V.~H., de la Cruz Rodr{\'{\i}}guez, J., \& Rouppe van der Voort, L.\ 2014, \apj, 781, 126


\bibitem[Pereira et al.(2015)]{Pereira:15} Pereira, T.~M.~D., Carlsson, M., De Pontieu, B., \& Hansteen, V.\ 2015, \apj, 806, 14 

\bibitem[Pesnell et al.(2012)]{Pesnell:12} Pesnell, W.~D., Thompson, B.~J., \& Chamberlin, P.~C.\ 2012, \solphys, 275, 3 

\bibitem[Peter et al.(2014)]{Peter:14} Peter, H., Tian, H., Curdt, W., et al.\ 2014, Science, 346, 1255726 


\bibitem[Rathore et al.(2015)]{Rathore:15} Rathore, B., Carlsson, M., Leenaarts, J., \& De Pontieu, B.\ 2015, \apj, 811, 81

\bibitem[Rouppe van der Voort et al.(2017)]{Luc:17} Rouppe van der Voort, L., De Pontieu, B., Scharmer, G.~B., et al.\ 2017, \apjl, 851, L6 


\bibitem[Scherrer et al.(2012)]{Scherrer:12} Scherrer, P.~H., Schou, J., Bush, R.~I., et al.\ 2012, \solphys, 275, 207 


\bibitem[Shelton et al.(2015)]{Shelton:15} Shelton, D., Harra, L., \& Green, L.\ 2015, \solphys, 290, 753 

\bibitem[Shibata et al.(1989)]{Shibata:89} Shibata, K., Tajima, T., Steinolfson, R.~S., \& Matsumoto, R.\ 1989, \apj, 345, 584

\bibitem[Shimizu(2015)]{Shimizu:15} Shimizu, T.\ 2015, Physics of Plasmas, 22, 101207


\bibitem[Smitha et al.(2018)]{Smitha:18} Smitha, H.~N., Chitta, L.~P., Wiegelmann, T., \& Solanki, S.~K.\ 2018, \aap, 617, A128 

\bibitem[Spadaro et al.(2004)]{Spadaro:04} Spadaro, D., Billotta, S., Contarino, L., Romano, P., \& Zuccarello, F.\ 2004, \aap, 425, 309 


\bibitem[Su et al.(2018)]{Su:18} Su, Y., Liu, R., Li, S., et al.\ 2018, \apj, 855, 77 


\bibitem[Tarr et al.(2014)]{Tarr:14} Tarr, L.~A., Longcope, D.~W., McKenzie, D.~E., \& Yoshimura, K.\ 2014, \solphys, 289, 3331 

\bibitem[Testa et al.(2016)]{Testa:16} Testa, P., De Pontieu, B., \& Hansteen, V.\ 2016, \apj, 827, 99

\bibitem[Tian et al.(2018)]{Tian:18} Tian, H., Zhu, X., Peter, H., et al.\ 2018, \apj, 854, 174 

\bibitem[Tian et al.(2016)]{Tian:16} Tian, H., Xu, Z., He, J., \& Madsen, C.\ 2016, \apj, 824, 96

\bibitem[Tian et al.(2014)]{Tian:14} Tian, H., Kleint, L., Peter, H., et al.\ 2014, \apjl, 790, L29

\bibitem[Toriumi et al.(2017)]{Toriumi:17} Toriumi, S., Katsukawa, Y., \& Cheung, M.~C.~M.\ 2017, \apj, 836, 63 


\bibitem[Vargas Dom{\'{\i}}nguez et al.(2014)]{Santiago:14} Vargas Dom{\'{\i}}nguez, S., Kosovichev, A., \& Yurchyshyn, V.\ 2014, \apj, 794, 140

\bibitem[Vargas Dom{\'{\i}}nguez et al.(2012)]{Santiago:12} Vargas Dom{\'{\i}}nguez, S., van Driel-Gesztelyi, L., \& Bellot Rubio, L.~R.\ 2012, \solphys, 278, 99

\bibitem[Vissers et al.(2015)]{Vissers:15} Vissers, G.~J.~M., Rouppe van der Voort, L.~H.~M., Rutten, R.~J., Carlsson, M., \& De Pontieu, B.\ 2015, \apj, 812, 11

\bibitem[Winebarger et al.(2013)]{Winebarger:13} Winebarger, A.~R., Walsh, R.~W., Moore, R., et al.\ 2013, \apj, 771, 21 

\bibitem[Yokoyama \& Shibata(1996)]{Yokoyama:96} Yokoyama, T., \& Shibata, K.\ 1996, \pasj, 48, 353 

\bibitem[Yokoyama \& Shibata(1995)]{Yokoyama:95} Yokoyama, T., \& Shibata, K.\ 1995, \nat, 375, 42 

\bibitem[Young et al.(2018)]{Young:18} Young, P.~R., Tian, H., Peter, H., et al.\ 2018, \ssr, 214, 120



\bibitem[Zhao et al.(2017)]{Zhao:17} Zhao, J., Schmieder, B., Li, H., et al.\ 2017, \apj, 836, 52 

\bibitem[Zuccarello et al.(2008)]{Zuccarello:08} Zuccarello, F., Battiato, V., Contarino, L., et al.\ 2008, \aap, 488, 1117

\bibitem[Zuccarello et al.(2005)]{Zuccarello:05} Zuccarello, F., Battiato, V., Contarino, L., et al.\ 2005, \aap, 442, 661 


\end{thebibliography}
\end{document}